\documentclass[cernpreprint, atlasdraft=true, texlive=2016, UKenglish, texmf]{atlasdoc}
 
\usepackage{atlaspackage}
\usepackage{atlasbiblatex}
 
\usepackage[BSM, jetetmiss, particle, process,snippets]{atlasphysics}
\newcommand*{\ttg}{\ensuremath{t\bar{t}\gamma}\xspace}
 
\usepackage{multirow}
\usepackage{siunitx}
\usepackage{rotating} 
\Crefname{figure}{Figure}{Figures}
\graphicspath{{logos/}{figures/}}
 
\usepackage{ANA-EXOT-2019-31-PAPER-defs}

\AtlasTitle{Observation of electroweak production of two jets in association with an isolated photon and missing transverse momentum, and search for a Higgs boson decaying into invisible particles at 13~$\text{TeV}$ with the ATLAS detector}
 
\AtlasAbstract{
This paper presents a measurement of the electroweak production of two jets in association with a $Z\gamma$ pair, with the $Z$ boson decaying into two neutrinos. It also presents a search for invisible or partially invisible decays of a Higgs boson with a mass of 125~$\text{GeV}$ produced through vector-boson fusion with a photon in the final state.
These results use data from LHC proton--proton collisions at $\sqrt{s}$ = 13~$\text{TeV}$ collected with the ATLAS detector and corresponding to an integrated luminosity of 139~fb$^{-1}$.
The event signature, shared by all benchmark processes considered for the measurements and searches, is characterized by a significant amount of unbalanced transverse momentum and a photon in the final state, in addition to a pair of forward jets. 
Electroweak $Z\gamma$ production in association with two jets is observed in this final state with a significance of 5.2 (5.1 expected) standard deviations. The measured fiducial cross-section for this process is 1.31$\pm$0.29 fb.
An observed (expected) upper limit of 0.37 ($0.34^{+0.15}_{-0.10}$) at 95\% confidence level is set on the branching ratio of a 125~$\text{GeV}$ Higgs boson to invisible particles, assuming the Standard Model production cross-section.
The signature is also interpreted in the context of decays of a Higgs boson into a photon and a dark photon. An observed (expected) 95\% CL upper limit on the branching ratio for this decay is set at 0.018 ($0.017^{+0.007}_{-0.005}$), assuming the Standard Model production cross-section for a 125~$\text{GeV}$ Higgs boson.
}
 
\author{The ATLAS Collaboration}
 
\AtlasJournalRef{\EPJC 82 (2022) 105}
\AtlasDOI{10.1140/epjc/s10052-021-09878-z}

\PreprintIdNumber{CERN-EP-2021-137}

\hypersetup{pdftitle={ATLAS document},pdfauthor={The ATLAS Collaboration}}
 
\begin{document}
 
\maketitle
 
\tableofcontents
\newpage
 
\section{Introduction}
\label{sec:intro}

Studying the self-couplings of the Standard Model (SM) vector bosons, precisely predicted through the SU(2)$_\textrm{L}\times$U(1)$_\textrm{Y}$ gauge symmetry, provides a unique opportunity to better understand the electroweak sector of the SM and gain insight into possible anomalies due to new phenomena.
Vector-boson scattering (VBS), $VV'\to~VV'$ with $V,V' = \Wboson/\Zboson/\gamma$, is an interesting process to study, being sensitive to both the triple and quartic gauge-boson couplings, which are particularly sensitive to the presence of physics effects beyond the Standard Model (BSM)~\cite{Eboli_2004,Eboli_2006,Degrande_2013}.
In hadron collisions, such as those produced at the Large Hadron Collider (LHC), VBS events occur whenever two vector bosons, radiated from the initial-state quarks, interact with each other~\cite{PhysRevD.35.1626}, producing a final-state signature characterized by the two vector bosons and a pair of forward hadronic jets in opposite hemispheres.
The two vector bosons can similarly annihilate and produce a SM Higgs boson in the so-called vector-boson fusion (VBF) mechanism.
Depending on the subsequent decay of the SM Higgs boson, different final-state signatures can be exploited to investigate its properties. Precise knowledge of the various Higgs boson decay branching ratios is a fundamental aspect of understanding whether the 125~\GeV{} scalar boson behaves according to the SM predictions or whether new physics phenomena modify the Higgs sector.
 
\begin{figure}[!h]
\centering
\subfloat[Strong diagram]{\includegraphics[width=0.25\textwidth]{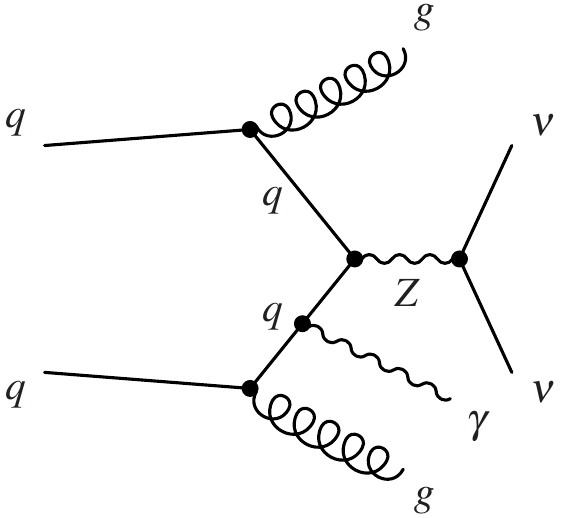}\label{fig:strong_zgamma}}
\hspace{1cm}
\subfloat[VBF electroweak diagram]{\includegraphics[width=0.25\textwidth]{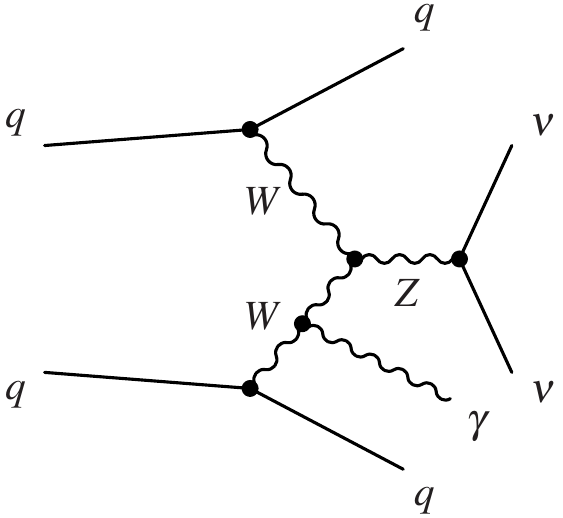}\label{fig:ewk_gamma}}
\hspace{1cm}
\subfloat[Triboson diagram]{\includegraphics[width=0.25\textwidth]{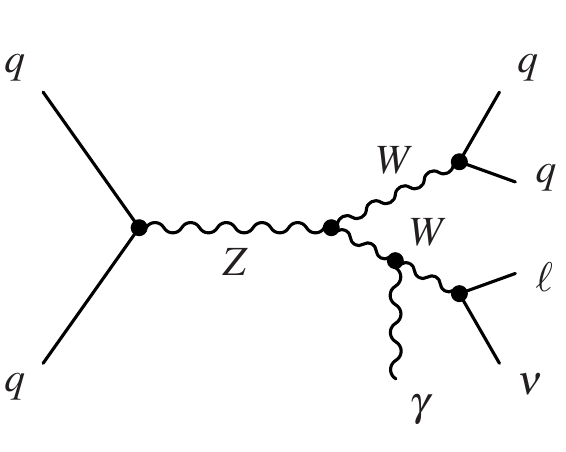}\label{fig:triboson_gamma}}
\caption{Representative Feynman diagrams for the dominant SM processes contributing to the considered signature:
\protect\subref{fig:strong_zgamma} is an example of the order $\alphas^2\alphaewk^3$ diagrams referred to as ``strong''; \protect\subref{fig:ewk_gamma} and \protect\subref{fig:triboson_gamma} are examples of the order $\alphaewk^5$ diagrams, which are collectively referred to as ``electroweak'' (EW).}
\label{fig:Feynman_strongbkg}
\end{figure}

This paper presents a set of SM measurements and searches for new phenomena in a final-state signature characterized by two forward hadronic jets, a photon, and a significant amount of unbalanced momentum in the plane transverse to the beam direction (\met) due to undetected particles.
In the SM this can result from \Vgamma production, where $V$ is either a \Zboson boson decaying into an undetected neutrino--antineutrino pair or a \Wboson boson decaying leptonically, where the charged lepton is not reconstructed in the detector. The latter is a background to the measurements and searches in this paper.
The $\Vgamma$ events are produced in the SM through a combination of `strong' and `electroweak' (EW) contributions: the former are produced through diagrams of order $\alphas^2\alphaewk{}^3$ at the Born level as shown in Figure~\ref{fig:strong_zgamma}, where \alphas{} is the strong coupling constant and \alphaewk{} is the electromagnetic coupling constant, while the latter are produced more rarely through diagrams of order $\alphaewk^5$ at the Born level\footnote{One order of \alphaewk{} is included for the decay of the vector boson.}  as shown in Figures~\ref{fig:ewk_gamma} and \ref{fig:triboson_gamma}.
The \ZgammaNoLabel cross-sections have been computed at next-to-leading order (NLO) in \alphas{} for both the strong~\cite{Campanario_2014} and EW~\cite{Campanario_2018} production modes.
It is not possible to study VBS diagrams, as shown in Figure~\ref{fig:ewk_gamma}, independently of other electroweak processes (e.g. triboson production as shown in Figure~\ref{fig:triboson_gamma}) as only the ensemble is gauge invariant~\cite{Accomando_2006}. There is also interference between the SM electroweak and strong processes, which is accounted for in the measurement.
 
In Run~1 the $\ZgammaNoLabel$ EW production cross-section has been measured in the dielectron and dimuon final states of the $Z$ boson by the ATLAS and CMS experiments~\cite{STDM-2017-26,CMS-SMP-14-018} with observed significances of 4.1 and 3.0 standard deviations, respectively. The CMS Collaboration measured EW \Zlljj{} after observing the process with a significance of 9.4 standard deviations in 137~\ifb{} of 13~\TeV{} proton--proton (\pp) collisions~\cite{cmscollaboration2021measurement}.
 
Profiting from the full 139~\ifb{} dataset collected with the ATLAS detector during Run~2 of the LHC, the analysis presented in this paper reports the observation of EW \Zgammajj{} production at the LHC.
 
The observation of EW production of \Zgammajj{} lays the groundwork for further investigation of this signature in looking for possible hints of BSM physics, based on interesting and well-motivated benchmark scenarios involving new dark matter (DM) candidate particles or a hidden sector of new particles coupling with the SM Higgs boson. The existence of DM is evident from astrophysical observations~\cite{ARBEY2021103865}, although its connection with the SM is still unknown because it has only been observed through gravitational interactions.
In this paper, the connection of DM with SM particles is explored by introducing a coupling with the 125~\GeV{} Higgs boson. A class of BSM scenarios, referred to as \emph{Higgs-portal} models~\cite{Lopez_Honorez_2012}, feature a DM candidate behaving as a singlet under the SM gauge symmetries, with the Higgs boson playing the role of a mediator between the DM candidate and SM particles.
 
Two main benchmark models are chosen when searching for new phenomena in the considered final-state signature, probing the invisible decay of the Higgs boson or its semi-visible decay into one invisible particle and one photon.
The decay of the Higgs boson into invisible particles (\Hinv{}) when produced through the SM-predicted VBF process in association with an emitted photon~\cite{Gabrielli:2007wf,Gabrielli:2016mdd} is probed in this paper for the first time. The Feynman diagram for this signature is shown in Figure~\ref{fig:Hinv_y}. Compared to a similar signature without the photon, this one benefits from better background rejection and a higher signal reconstruction efficiency; however, it has a lower production cross-section.
 
No direct constraints on the \Hinv{} decays in this experimental signature currently exist.
Constraints on the invisible Higgs boson branching ratio (\Binv) were set by the ATLAS Collaboration using the full Run~1 dataset at 7 and 8~\TeV~\cite{HIGG-2013-03,HIGG-2013-16, HIGG-2014-07} and the full Run~2 dataset at 13~\TeV~\cite{EXOT-2016-37, HIGG-2018-54}. Similar searches were carried out by the CMS Collaboration~\cite{CMS-HIG-13-030,CMS-HIG-16-016,CMS-HIG-17-023} using both the Run~1 and Run~2 datasets.
The most stringent limits are from the statistical combination of the search results, for which ATLAS reports an observed (expected) upper limit on \Binv of 0.26 (0.17) and CMS reports an upper limit of 0.19 (0.15), all at 95\% confidence level (CL). In these combinations, the VBF production channel is the single channel with the highest expected sensitivity, for which ATLAS and CMS reported observed (expected) 95\% CL limits on \Binv{} of 0.37 (0.28) and 0.33 (0.25), respectively, using 36~\ifb of Run~2 data.
 
In addition to the \Hinv{} benchmark, this analysis probes a dark-photon model~\cite{Gabrielli_2014, Gabrielli:2014oya, Biswas:2016jsh} which predicts a light or massless\footnote{For the results in this paper, the signal acceptance changes by less than 1\% for dark-photon masses up to 10~\GeV{}.} dark photon ($\gammad$) coupled with the Higgs boson through a U(1) unbroken dark sector. The Feynman diagram for this signature is shown in Figure~\ref{fig:Hyyd}, and BSM extensions to other resonance masses are possible~\cite{Tsai_2017}. In this model, a Higgs boson decays into a photon and an invisible dark photon with a branching ratio \Byyd. CMS has probed this benchmark model by considering associated $ZH$ production and VBF Higgs boson production~\cite{CMS-EXO-19-007,CMS-EXO-20-005}, and reported observed (expected) 95\% CL upper limits of 0.046 (0.036) and 0.035 (0.028) on \Byyd in their respective signatures.  These two results were combined to yield an observed (expected) 95\% CL upper limit of 0.029 (0.021).
 
\begin{figure}[!h]
\centering
\subfloat[Invisible Higgs boson signal]{\includegraphics[width=0.25\textwidth]{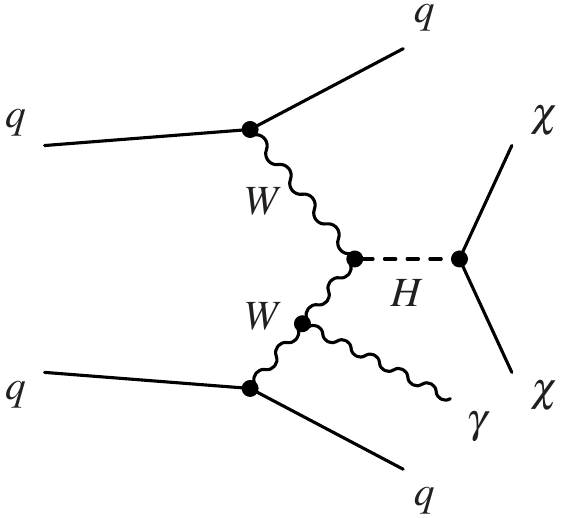}\label{fig:Hinv_y}}
\hspace{1cm}
\subfloat[Dark-photon signal]{\includegraphics[width=0.25\textwidth]{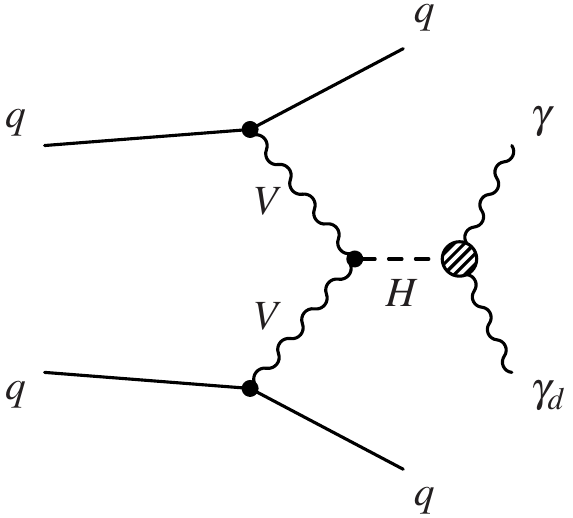}\label{fig:Hyyd}}
\caption{Representative Feynman diagrams corresponding to the dominant BSM signal processes:
\protect\subref{fig:Hinv_y} the signal due to the invisible Higgs boson decay produced through VBF in association with an emitted photon. The incoming quarks appearing on the left hand side of the diagram are different from each other. \protect\subref{fig:Hyyd} the signal due to the decay of the Higgs boson to a photon--dark-photon pair, produced through VBF.}
\label{fig:signals}
\end{figure}
 
The SM processes resulting in the same final-state signature as described before, mainly \Zgamma and \Wgamma, represent background contributions in the searches for new phenomena. These processes were simulated, for both the \ZgammaNoLabel EW cross-section measurement and the searches for new phenomena, through dedicated Monte Carlo samples, detailed in \cref{sec:simulation}. Minor and reducible background contributions arise due to the misreconstruction of objects in the detector, including hadronic jets reconstructed as photons. These background processes are estimated using simulation and data-driven techniques and are validated using dedicated data samples.
The SM $H\to \Zboson(\to \nu\nu)\gamma$ process produces the same signature as the investigated $\Hinv+\gamma$ signal, but the small contribution of such events satisfying the criteria defined in \cref{sec:event_selection} is neglected in the searches for new phenomena.
 
A brief description of the ATLAS detector is provided in Section~\ref{sec:detector}.
The dataset, simulated samples, and physics object selection are covered in Sections~\ref{sec:data},~\ref{sec:simulation}, and~\ref{sec:objects}, respectively. The analysis strategies and the event selection for the EW $\Zgammajj$ measurement and searches based on different signal hypotheses, \Hinv{} and \Hyyd{}, are discussed in Section~\ref{sec:event_selection}. The modelling of the different processes contributing to the considered signature is presented in Section~\ref{sec:analysis}, and the fit models and extracted results for the different signal hypotheses are described in Section~\ref{sec:fit_models}.

\section{ATLAS detector}
\label{sec:detector}

\newcommand{\AtlasCoordFootnote}{
ATLAS uses a right-handed coordinate system with its origin at the nominal interaction point (IP)
in the centre of the detector and the \(z\)-axis along the beam pipe.
The \(x\)-axis points from the IP to the centre of the LHC ring,
and the \(y\)-axis points upwards.
Cylindrical coordinates \((r,\phi)\) are used in the transverse plane,
\(\phi\) being the azimuthal angle around the \(z\)-axis.
The pseudorapidity is defined in terms of the polar angle \(\theta\) as \(\eta = -\ln \tan(\theta/2)\).
Angular distance is measured in units of \(\DeltaR \equiv \sqrt{(\Delta\eta)^{2} + (\Delta\phi)^{2}}\).}

The ATLAS experiment~\cite{PERF-2007-01,ATLAS-TDR-19,PIX-2018-001} at the LHC is a multipurpose particle detector
with a forward--backward symmetric cylindrical geometry and near \(4\pi\) coverage in
solid angle.\footnote{\AtlasCoordFootnote}
It consists of an inner tracking detector surrounded by a thin superconducting solenoid
providing a \SI{2}{\tesla} axial magnetic field, electromagnetic and hadron calorimeters, and a muon spectrometer.
The inner tracking detector (ID) covers the pseudorapidity range \(|\eta| < 2.5\).
It consists of silicon pixel, silicon microstrip, and transition radiation tracking detectors.
Lead/liquid-argon (LAr) sampling calorimeters provide electromagnetic (EM) energy measurements
with high granularity.
A steel/scintillator-tile hadron calorimeter covers the central pseudorapidity range (\(|\eta| < 1.7\)).
The endcap and forward regions are instrumented with LAr calorimeters
for both the EM and hadronic energy measurements up to \(|\eta| = 4.9\).
The muon spectrometer (MS) surrounds the calorimeters and is based on
three large superconducting air-core toroidal magnets with eight coils each.
The field integral of the toroids ranges between \num{2.0} and \SI{6.0}{\tesla\metre}
across most of the detector.
The muon spectrometer includes a system of precision tracking chambers and fast detectors for triggering.
A two-level trigger system is used to select events.
The first-level (L1) trigger is implemented in hardware and uses a subset of the detector information
to accept events at a rate below \SI{100}{\kHz}.
This is followed by a software-based high-level trigger (HLT)~\cite{TRIG-2016-01,TRIG-2019-04} which
reduces the accepted event rate to \SI{1}{\kHz} on average
depending on the data-taking conditions. An extensive software suite~\cite{ATL-SOFT-PUB-2021-001} is used in the reconstruction and analysis of real and simulated data, in detector operations, and in the trigger and data acquisition systems of the experiment.

\section{Data samples}
\label{sec:data}

This analysis relies on the data collected by the ATLAS experiment from LHC \pp collisions at \rts~=~13~\TeV{} during 2015--2018 stable beam conditions when all subdetectors were operational~\cite{DAPR-2018-01}, corresponding to a total integrated luminosity of \integLumi~\cite{ATLAS-CONF-2019-021,LUCID2}.
The data used in this analysis were recorded mainly using trigger algorithms based on the presence of missing transverse momentum, \met{} (described in Section \ref{sec:objects})~\cite{TRIG-2019-01}. The trigger thresholds for the \met{} were determined by the data-taking conditions during the different periods, especially by the average number of multiple \pp interactions in the same or neighbouring bunch crossings, referred to as \pileup. The first-level trigger threshold was 50--55~\GeV{}, depending on the data-taking period, and the lowest-transverse-momentum (\pT{}) unprescaled HLT threshold for the \met{} trigger algorithm was 70~\GeV{} in 2015, 90~\GeV{} in 2016 and 110~\GeV{} in 2017--2018. This corresponds to a trigger operating on its maximum-efficiency plateau for events with offline \met{} of $\sim$180~\GeV{}, depending on the trigger thresholds, in this final state.
Independent data samples exploited to study \Wgamma background were collected using the lowest-\pT{} unprescaled single-lepton triggers, with \pT{} thresholds ranging from 20 to 26~\GeV{} for the triggers with the tightest lepton identification criteria~\cite{TRIG-2018-01,TRIG-2018-05}.
The \met triggers complemented the muon triggers to increase the number of accepted single-muon events by 28\%, as discussed in Section~\ref{subsec:crdef}.

\section{Simulated event samples}
\label{sec:simulation}


Monte Carlo (MC) simulated samples are used to model both the SM and BSM processes. 
The full set of simulated samples is summarized in
Table~\ref{tab:samples}. The generated events were processed through a
simulation~\cite{SOFT-2010-01} of the ATLAS detector geometry and
response using \GEANT~\cite{Agostinelli:2002hh}, and through the same
reconstruction software as the collected data.
For BSM signal samples with a Higgs boson mass different from 125~\GeV{}, the detector response was simulated using a fast parameterized simulation of the ATLAS calorimeters~\cite{ATL-SOFT-PUB-2018-002} and the full \GEANT{} simulation for the other subdetectors.
 
The \pileup effects in the same and neighbouring proton-bunch
crossings were modelled by adding detector signals from simulated inelastic \pp events to the original hard-scattering (HS) event. These were generated with \PYTHIA[8.186]~\cite{Sjostrand:2007gs} using the \NNPDF[2.3lo] set of parton distribution functions (PDFs)~\cite{Ball:2012cx} and the A3 set of tuned parameters (tune)~\cite{ATL-PHYS-PUB-2016-017}.
The energy scale and resolution for leptons and jets, their reconstruction and identification efficiencies, and the trigger efficiencies in the simulation are corrected to match those measured in data.
 
\begin{table}[!htbp]
\caption{Summary of generators used for simulation. The details and
the corresponding references are provided in the body of the
text. The $V$ in \Vjet{} represents either a $W$ or a $Z$ boson. The calculation precision indicates the \alphas order of the expansion.}
\label{tab:samples}
\centering
\resizebox{\textwidth}{!}{
\def\arraystretch{1.5}
\newcolumntype{L}[1]{>{\raggedright\let\newline\\\arraybackslash}p{#1}}
\begin{tabular}{L{0.175\textwidth}L{0.29\textwidth}L{0.18\textwidth}L{0.19\textwidth}L{0.16\textwidth}L{0.1\textwidth}}
\toprule
Process & Generator & ME Order & PDF & Parton Shower & Tune \\
\midrule
\multicolumn{6}{c}{SM process samples}\\
\midrule
Strong \Vg{} & \SHERPA[2.2.8] & \NLO (up to 1-jet), \newline LO (up to 3-jets) & \NNPDF[3.0nnlo] & \Sherpa \MEPSatNLO & \Sherpa \\
EW \Vg{} & \MGNLO[2.6.5] & LO & \NNPDF[3.1lo] & \PYTHIA[8.240] &   A14 \\
EW $VV\textrm{+\,jets}$ & \Sherpa[2.2.1] or \Sherpa[v2.2.2] & LO & \NNPDF[3.0nnlo] & \Sherpa \MEPSatNLO & \Sherpa \\
$VV\textrm{+\,jets}$ & \Sherpa[2.2.1] or \Sherpa[2.2.2] & \NLO (up to 1-jet), LO (up to 3-jets)
& \NNPDF[3.0nnlo] & \sherpa \MEPSatNLO{} & \Sherpa \\ 
EW \Vjet{} & \HERWIG[7.1.3] or \HERWIG[7.2.0] & \NLO & \MMHT2014nlo68cl
&
\HERWIG[7.1.3] & \HERWIG[7] \\ 
Strong\newline \Wmn{}/\newline \Wtn{} & \Sherpa[2.2.7] & \NLO (up to 2-jets), \newline LO (up to 4-jets)
& \NNPDF[3.0nnlo] & \Sherpa \MEPSatNLO & \Sherpa \\
\ttg{} & \MGNLO[2.2.3] & \NLO & \NNPDF[2.3lo] & \PYTHIA[8.186] & A14 \\
\ttbar/$Wt$ & \POWHEGBOX[v2] & \NLO & \NNPDF[3.0nlo] & \PYTHIA[8.230] &A14 \\
$V\gamma\gamma$ & \Sherpa[2.2.2] (at 0-jet), LO (up to 2-jets) & \NLO & \NNPDF[3.0nnlo] & \sherpa\MEPSatNLO{} & \Sherpa \\
\gjet{} & \Sherpa[2.2.2] & \NLO (up to 2-jets), \newline LO (up to 4-jets) & \NNPDF[3.0nnlo] & \Sherpa \MEPSatNLO & \Sherpa \\
\midrule
\multicolumn{6}{c}{Higgs-related samples}\\
\midrule
ggF Higgs & \POWHEG[v2~NNLOPS] & \NNLO & \PDFforLHC[15] & \PYTHIA[8.230] & AZNLO \\ 
Higgs\,+\,$\gamma$ & \MGNLO[2.6.2] & \NLO & \PDFforLHC[15]  & \HERWIG[7.1.3p1] & A14 \\
ggF Higgs$\to\gamma\gammad$ & \POWHEG[v2~NNLOPS] & NNLO & \PDFforLHC[15]  & \PYTHIA[8.244p3] & AZNLO \\ 
VBF Higgs$\to\gamma\gammad$ & \POWHEG[v2] & \NLO & \CTEQ[6L1] & \PYTHIA[8.244p3] & AZNLO \\
\midrule
\multicolumn{6}{c}{Systematic variation samples}\\
\midrule
\Vg{} $\alphaewk^4$ interference & \MGNLO[2.6.2] & LO & \NNPDF[3.1lo] & \PYTHIA[8.240] & AZNLO \\
 
\bottomrule
\end{tabular}
}
\end{table}

\subsection{\Vg{} processes}
\label{subsec:Vggeneration}
The \Wgamma{} and \ZgammaNoLabel{} (together labelled \Vg{}) processes contributing to the signature considered in this analysis contain a charged lepton ($\ell$ = $e$, $\mu$ or $\tau$) and a neutrino, a pair of neutrinos ($\nu\nu$) or a pair of charged leptons ($\ell\ell$) together with a photon and associated jets. The \Vg{} processes are split into two components based on the order in the electroweak coupling constant \alphaewk{}. At tree level, the strong component is of order $\alphas^2\alphaewk{}^3$ and the EW component is of order $\alphaewk{}^5$; example Feynman diagrams are shown in \cref{fig:Feynman_strongbkg} for these different contributions.
The strong component of each \Vg{} contribution was simulated for photon \pT{} greater than 7~\GeV{} using the \SHERPA[2.2.8]~\cite{Bothmann:2019yzt} event generator at \NLO precision in \alphas for up to one additional parton and at \LO precision in \alphas for up to three additional partons. These calculations use the Comix~\cite{Gleisberg:2008fv} and OpenLoops~\cite{Cascioli:2011va} matrix element generators, and the parton-shower matching~\cite{Schumann:2007mg} was performed using the \mepsatlo{}~\cite{Hoeche:2012yf} or \mepsatnlo~\cite{Hoeche:2011fd,Hoeche:2012yf,Catani:2001cc,Hoeche:2009rj} prescription. The \NNPDF[3.0nnlo]{} set of PDFs~\cite{Ball:2014uwa} was used, along with dedicated parton shower tuning developed by the \SHERPA authors. Electroweak radiative corrections to strong \Vg{} production have been computed at NLO~\cite{Denner_2016,Hollik_2004,ZgNLOEWK} as weights in \SHERPA, and these are roughly $-2$\% to $-4$\% relative corrections in the chosen signal region.
A second sample of \Zgamma{} was generated using \MGNLO~\cite{Alwall:2014hca} with the \FXFX merging scheme~\cite{Frederix:2012ps} at \NLO precision in \alphas for up to one additional parton, filtered for photon \pT$>$10~\GeV, and showered using \PYTHIA[8]~\cite{Sjostrand:2014zea}. The difference between the \SHERPA and \MGNLO \Zgamma{} predictions is symmetrized around the \SHERPA{} prediction and is taken as a source of modelling systematic uncertainty.
 
Matrix elements for the EW contribution were calculated at \LO{} in \alphas{} using \MGNLO[2.6.5], and the photon \pT{} was required to be larger than 10~\GeV{}. Generated events were showered using \PYTHIA[8]~\cite{Sjostrand:2014zea} with the dipole recoil option enabled along with the CKKW-L~\cite{Lonnblad:2001iq,Lonnblad:2011xx} merging scheme.
These EW samples are normalized to \NLO QCD predictions obtained from \VBFNLO~\cite{VBFNLO} through a correction which depends on the dijet invariant mass (\mjj{}).
The effect of the QCD factorization and renormalization scale choices is evaluated from the same \NLO QCD process calculation in \VBFNLO.
The \MGNLO[2.6.5] EW samples include VBS contributions (shown in \cref{fig:ewk_gamma}) and triboson contributions (shown in \cref{fig:triboson_gamma}) which can depend on trilinear and quartic gauge couplings.
The triboson processes contribute as much as 15\% of the EW samples for low $\mjj$ values, $250 < \mjj < 500$~\GeV, but less than 3\% in the more signal-like high-\mjj region.
The interference between the EW and strong-production diagrams, which is of order $\alphas\alphaewk{}^4$, was simulated at LO in \alphas{} in \MGNLO[2.6.5] and the corresponding events were showered using \PYTHIA[8] with the dipole recoil shower enabled. This simulated sample is not included in the background predictions but is used to calculate an uncertainty in the EW \Vg{} contribution, which is 5\% or less in the $\mjj{} > 500$~\GeV{} kinematic region. Similarly, the EW production of a $W$ or $Z$ boson in association with two photons was found to be less than 2\% of the EW \Vg{} samples calculated with \MGNLO[2.6.5], so it is not included.
 
The whole \Vg{} simulation (i.e. both EW and strong production) applies smooth-cone isolation~\cite{Frixione_1998} with a cone size of 0.1 and parameters $n = 2$ and $\epsilon = 0.10$ to remove the collinear singularity between photons and charged partons which would otherwise appear in the process amplitude calculations.
 
\subsection{\Vjet{} processes}
\label{subsec:Vjgeneration}
 
Simulation is used to model \Vjet{} processes when the vector boson decays into neutrinos, muons or $\tau$-leptons, and most of these simulated events are removed by the lepton vetoes, object overlap removal described in \cref{sec:objects}, and the removal of overlaps with \Vg{}, which is described later in this section. Therefore, their contribution is very small. However, $V$ decays into electrons enter the selection mostly through the electrons being misidentified as photons, and this is modelled using the fully data-driven method described in~\cref{subsec:background}.
The strong-production \Vjet{} sample was simulated with the \SHERPA[2.2.7] event generator with the \NNPDF[3.0nnlo] set of PDFs~\cite{Ball:2014uwa}.
Parton-shower matching~\cite{Schumann:2007mg} was performed using either the \MEPSatLO{} or \MEPSatNLO{} prescription with associated parameters tuned by the \SHERPA authors. The strong production of \Vjet{} uses \NLO matrix elements for up to two partons and \LO matrix elements for up to four partons calculated with the Comix and OpenLoops libraries and the \mepsatnlo prescription. The samples are normalized to a next-to-next-to-leading-order (\NNLO) prediction for \Vjet~\cite{Anastasiou:2003ds}.
The EW \Vjet{} sample was generated at \NLO in \alphas{}, using \herwig7~\cite{Bellm:2015jjp} to perform the parton shower and hadronization with the \MMHT2014nlo68cl PDF set~\cite{Harland-Lang:2014zoa}. \herwig7{} uses its Matchbox module~\cite{Matchbox} to assemble calculations with dipole shower algorithms~\cite{DipoleShower} and interfaces with matrix element plug-ins from \VBFNLO{} to compute NLO \alphas{} corrections in the VBF approximation.
 
After generating events for both the strong and EW \Vg{} processes as in \cref{subsec:Vggeneration}, the samples may overlap with \Vjet{} events in which an ISR/FSR photon is radiated. To avoid this overlap, \Vg{} events are removed if the photon passes the smooth-cone isolation and lies within $\Delta R=0.1$ of an electron, muon, or $\tau$-lepton that has $\pT>10$~\GeV{}. In \Vjet{} samples, events are accepted only if they satisfy the same criteria.
 
\subsection{Top-quark processes}
 
Minor background contributions originating from top-quark-related processes and associated production of top quarks with a \Wboson{} boson, were all modelled using the \POWHEGBOX[v2]~\cite{Nason:2004rx,Frixione:2007vw,Alioli:2010xd} generator at \NLO{} with the \NNPDF[3.0nlo] PDF set. The events were interfaced with \PYTHIA[8.230] for the parton shower and hadronization modelling with the A14 tune~\cite{ATL-PHYS-PUB-2014-021} and the \NNPDF[3.0nlo] set of PDFs. The production of \ttg{} was modelled using the \MGNLO[2.2.3] generator at \NLO{}. The events were interfaced with \PYTHIA[8.186] for the parton shower and hadronization modelling with the A14 tune and the \NNPDF[2.3lo] set of PDFs. The decays of bottom and charm hadrons were performed by \EVTGEN[1.6.0]~\cite{Lange:2001uf} in all top-quark processes.
 
\subsection{Additional background samples}
\label{subsection:misrecobkgs}
 
The \gjet{} background is simulated using \SHERPA[2.2.2] at \NLO{} in \alphas{}, similarly to the \SHERPA{} \Vjet{} samples in Section~\ref{subsec:Vggeneration} except for the use of a dynamical merging scale to capture the fragmentation contribution~\cite{Siegert:2016bre}. Due to the large cross-section for this process, a partially data-driven `jet-smearing' technique was used to increase the sample size as discussed in \cref{subsec:background}.
 
Other processes listed in Table~\ref{tab:samples} but not described in the previous subsections result in a negligible contribution to the analysis signature. Therefore, no further description is given.

\subsection{Higgs boson processes}
\label{sec:sig_process}
In the searches for \Hinv{} carried out in this paper the target is VBF Higgs boson production, although the small contribution from gluon--gluon fusion (ggF) Higgs boson production processes satisfying the selection criteria is also considered as a contributing signal.
For both of these processes, corresponding MC samples were generated according to the details described in the following.
The invisible Higgs boson decay was simulated using the SM $H\to ZZ^{*}\to 4\nu$ decay with a 100\% branching ratio.
The difference in the relevant kinematic distributions between the $H\to 4\nu$ process and Higgs boson decays into new undetected particles is negligible.
 
The VBF Higgs boson production process with an additional photon, as shown in \cref{fig:Hinv_y}, was simulated for a Higgs boson mass of 125 \GeV{}, requiring the presence of only electroweak vertices at \LO{} and photon \pT{} larger than 10~\GeV{}. This process was computed to \NLO accuracy in \alphas using \MGNLO interfaced with \HERWIG[7]~\cite{Bahr:2008pv,Bellm:2015jjp} for parton shower and non-perturbative hadronization effects, using the \PDFforLHC[15] PDF set~\cite{Butterworth:2015oua}. Parton shower uncertainties are computed from the relative difference between the process generated by \MGNLO at \LO in \alphas with showering by \HERWIG[7] and the same process with showering by \PYTHIA[8]~\cite{Skands:2014pea} with the dipole recoil shower variable~\cite{Cabouat_2018} turned on. The parton shower comparison is not made at \NLO{} because the dipole recoil option in \PYTHIA[8] is available only at \LO{}, and without it there is a larger prediction of central emissions.\footnote{The option gives an alternative approach to local recoils, where only one final-state parton takes the recoil of an emission. The dipole recoil option has been shown to better model the CMS data in VBF \Zll{} processes~\cite{CMS-SMP-16-018,ATL-PHYS-PUB-2019-004}. When it is used, it improves the modelling of radiative emissions from non-colour-connected partons (the VBF jets), which are poorly modelled by \PYTHIA[8] when it is turned off.} The same smooth-cone isolation as described in Section~\ref{subsec:Vggeneration} is used to remove the collinear singularity between photons and charged partons. 
 
The ggF Higgs boson production process was simulated at \NNLO accuracy in \alphas using \POWHEG NNLOPS~\cite{Nason:2004rx,Frixione:2007vw,Alioli:2010xd,Hamilton:2013fea,Hamilton:2015nsa}, which achieves NNLO accuracy for arbitrary inclusive $gg\to H$ observables by reweighting the Higgs boson rapidity spectrum in MJ-\MINLO~\cite{Hamilton:2012np,Campbell:2012am,Hamilton:2012rf} to that in HNNLO~\cite{Catani:2007vq}. The \PDFforLHC[15] PDF set~\cite{Butterworth:2015oua} and the \AZNLO tune of \PYTHIA[8] were used. This simulation was interfaced with \PYTHIA[8] for parton shower and non-perturbative hadronization effects.
The ggF prediction from the MC samples is normalized to the next-to-\NNLO cross-section in QCD plus electroweak corrections at \NLO~\cite{deFlorian:2016spz,Anastasiou:2016cez,Anastasiou:2015ema,Dulat:2018rbf,Harlander:2009mq,Harlander:2009bw,Harlander:2009my,Pak:2009dg,Actis:2008ug,Actis:2008ts,Bonetti:2018ukf}.
Photons present in these samples were generated using \textsc{Photos}~\cite{Golonka:2005pn,Davidson:2010ew} since no complete matrix element computation of ggF Higgs boson with a photon is available.\footnote{The \POWHEG{} sample does not include photon radiation from charged particles running in the Higgs boson production loop.}

The simulation of ggF production of a 125~\GeV{} Higgs boson described above is also used to model its decay into a photon and an invisible massless dark photon (\gammad) again normalized to a 100\% branching ratio. The VBF production process was simulated to \NLO precision in \alphas with the \POWHEG generator interfaced with \PYTHIA[8] for hadronization and showering and the same $\gamma\gammad$ decay, as shown in \cref{fig:Hyyd}. The \NLO{} electroweak corrections for VBF Higgs boson production were computed using HAWK~\cite{Denner:2014cla} and were applied as a function of the Higgs boson's \pT{}. The VBF samples were generated not only for a 125~\GeV{} Higgs boson, but also for lighter and heavier Higgs bosons, for the interpretation of the search in the context of other scalar mediators. Throughout this paper, the samples with a Higgs boson assume SM couplings and use the narrow width approximation~\cite{deFlorian:2016spz} for the various Higgs boson masses, ranging from 60~\GeV{} to 2~\TeV.



\section{Object reconstruction}
\label{sec:objects}

Objects are reconstructed from detector signatures by using identification algorithms widely deployed in ATLAS analyses.
Candidate events are required to have a reconstructed vertex with at least two associated tracks, each with $\pT > 0.5\ \GeV{}$ and originating from the beam collision region in the $x$--$y$ plane. The primary vertex in the event is selected as the vertex with the highest scalar sum of the squared \pT{} of associated tracks~\cite{ATL-PHYS-PUB-2015-026}.
 
Electrons are reconstructed by matching clustered energy deposits in the EM calorimeters to tracks in the ID~\cite{EGAM-2018-01}, including the transition regions between the barrel and endcap EM calorimeters at $1.37 < \abseta < 1.52$. Electron candidates must have $\pT > 4.5~\GeV{}$ and $\abseta < 2.47$, and fulfil loose identification criteria.
Depending on the \pT and \abseta range, muons are reconstructed by matching ID tracks to MS tracks or track segments, by matching ID tracks to a calorimeter energy deposit compatible with a minimum-ionizing particle, or by identifying MS tracks passing a loose requirement and compatible with originating from the IP~\cite{MUON-2018-03}. Muon candidates must have $\pT > 4~\GeV{}$ and $\abseta < 2.7$, and fulfil a very loose identification criterion.
No isolation requirement is placed on electron or muon candidates used as a lepton veto.
In events with one or more leptons associated with the primary vertex, i.e. in control regions described in Section~\ref{sec:event_selection}, muons (electrons) are required to satisfy medium (tight) identification criteria in order to improve the purity of background processes.
 
Photon candidates are reconstructed from clustered energy deposits in the EM calorimeter~\cite{EGAM-2018-01}. They must have $\pT > 15~\GeV{}$, fulfil tight identification and isolation criteria~\cite{EGAM-2018-01}, and lie within $\abseta < 2.37$ but not in the transition region ($1.37 < \abseta < 1.52$) between barrel and endcap EM calorimeters.
 
Particle flow (PFlow) jets are reconstructed using the \akt algorithm~\cite{Cacciari:2008gp,Fastjet} with a radius parameter of $R = 0.4$, using charged constituents associated with the primary vertex and neutral PFlow constituents as inputs~\cite{PERF-2015-09}. The jet energy is calibrated with the effect of \pileup removed~\cite{JETM-2018-05}. 
Jets are required to have $\pT > 20~\GeV{}$ and $\abseta < 4.5$. For jets with $\pT < 60~\GeV{}$ and $\abseta < 2.5$ the jet vertex tagger (JVT) discriminant~\cite{ATLAS-CONF-2014-018} is used to identify jets originating from the HS interaction through the use of tracking and vertexing. The chosen JVT working point corresponds to a selection efficiency for HS jets of about 97\%, evaluated on an inclusive $\Zboson(\to\mu\mu)$ + jets sample. For $\abseta > 2.5$, the two leading jets must pass a forward jet vertex tagger algorithm (fJVT)~\cite{ATL-PHYS-PUB-2015-034,ATL-PHYS-PUB-2019-026} that accepts 93\% of jets from the HS interaction and rejects about 58\% of \pileup jets with $\pt > 50~\GeV$, evaluated on an inclusive $\Zboson(\to\mu\mu)$ + jets sample.
Jets containing $b$-hadrons (\bjets) are identified using a multivariate discriminant (MV2c10) output distribution~\cite{PERF-2016-05}. The working point is chosen to provide a 77\% \bjet efficiency on an inclusive \ttbar sample, with rejection factors of 6 and 134 for charm-hadron jets and light-flavour quark- or gluon-initiated jets, respectively.
 
To avoid double counting of energy deposits, the reconstructed objects are required to be separated according to the procedure detailed in Table~\ref{tab:overlap_removal}. The overlap of photons with electrons, muons and jets is resolved by a \DeltaR{} criterion. For leptons in the vicinity of jets, the \DeltaR{} threshold for sufficient separation depends on the \pT{} of the lepton to account for the collimation of boosted objects.

\begin{table}[!htbp]
\caption{Overview of the overlap removal between objects and the corresponding matching criteria, listed according to decreasing priority.}
\label{tab:overlap_removal}
\centering
\begin{tabular}{ccc}
\toprule
Remove & Keep & Matching criteria \\
\hline
electron & electron & shared inner-detector track, electron with lower \pt removed \\
muon & electron & muon with calorimeter deposits and shared inner-detector track \\
electron & muon & shared inner-detector track \\
photon & electron & $\DeltaR < 0.4$ \\
photon & muon & $\DeltaR < 0.4$ \\
jet & electron & $\DeltaR < 0.2$ \\
electron & jet & $\DeltaR < \mathrm{min}(0.4, 0.04 + 10~\GeV/\pT^e)$ \\
jet & muon & number of tracks $<$ 3 and $\DeltaR < 0.2$ \\
muon & jet & $\DeltaR < \mathrm{min}(0.4, 0.04 + 10~\GeV/\pT^{\mu})$ \\
jet & photon & $\DeltaR < 0.4$ \\
\bottomrule
\end{tabular}
\end{table}
 
The unbalanced momentum in the transverse plane, referred to as missing transverse momentum or \metvec{}, is defined as the negative vectorial sum of the transverse momenta of all selected electrons, muons, photons, and jets, as well as tracks compatible with the primary vertex but not matched to any of those objects, this last contribution being called the \emph{soft term}~\cite{PERF-2016-07,ATLAS-CONF-2018-023}.
To define an estimate of the boson \pt in decays with charged leptons, events containing one or more selected leptons have the \metvec{} evaluation modified by treating such leptons as invisible particles. Each prompt lepton $\vec{p}_{\text{T}}$ is vectorially added to the \metvec to define the quantity \metnolepvec, as described in Ref.~\cite{ATLAS-CONF-2018-023}. The magnitude of the \metvec{} (\metnolepvec{}) is denoted by \met (\metnolep{}).
A related event property \etjetnojvt{} is the magnitude of the negative vectorial sum of all jets in the event with $\pt > 20~\GeV{}$ before the JVT requirement. This is a powerful variable for rejecting events where the \met is generated by inefficiencies of the JVT requirement for HS jets.
 
Several cleaning requirements are applied to suppress non-collision backgrounds~\cite{ATLAS-CONF-2015-029}. Misreconstructed jets can be caused by electronic noise, and jets from collisions are identified by requiring a good fit to the expected pulse shape for each constituent calorimeter cell. Beam-halo interactions with the LHC collimators are another source of misreconstructed jets. Those jets are identified by requirements on their energy distribution in the calorimeter and the fraction of their constituent tracks that originate from the primary vertex. The event is rejected if any selected jet is identified as a misreconstructed jet. Residual contributions of non-collision jets are absorbed into the normalization for the \gjet{} background as described in Section~\ref{subsec:background}.
 

\section{Event selection}
\label{sec:event_selection}

Based on the reconstructed objects in each event final state, the collected data are assigned to disjoint samples identified with specific phase-space regions, used in this analysis for different purposes.
The signal regions (SRs) have the highest purity of signal process events, with mostly irreducible background contributions.
The control regions (CR) are enriched in background processes and are used to constrain estimates of such contributions.
The validation regions (VR), similar in background process content to the CRs, are used to quantify the level of agreement between data and background predicted yields but are not included in the fit.
 
A baseline set of requirements is applied to select events with a photon, high \MET{}, and the VBF jet signature considered in this paper, as described in \cref{subsec:srdef}.
For the EW \Zg{} cross-section measurement, events satisfying the baseline SR requirements are categorized according to their dijet invariant mass \mjj, as described in Section~\ref{subsec:mjjbinning}. In the same section the fiducial volume considered for the cross-section measurement is defined.
The SR-selected events for the \Hinv{} search are split into categories of different signal purities based on a multivariate analysis discriminant, which is described in Section~\ref{subsec:keras}. The SR selection and fitted discriminant are modified for the resonant kinematics of the \Hyyd{} search, which are described in Section~\ref{subsec:mtbinning}.
In the statistical analysis of the data, described in \cref{sec:fit_models}, CR events are classified according to the same strategy described for the signal regions in the aforementioned measurement and searches.
 
\begin{table}[!htbp]
\caption{Summary of the requirements defining the baseline SR and various CRs considered in this analysis. Where present, the values in square brackets refer to the regions defined in the search for a \Hyyd signal. The leading and subleading jets must satisfy the fJVT requirements mentioned in Section~\protect\ref{sec:objects}. In the SR, \ZrcCR, and Low-\met VR definitions $\metnolep\equiv\met$ since no lepton is present. The \MT{} variable is defined in \cref{subsec:mtbinning}. When the same requirement is applied to multiple regions, this is reported once in the corresponding row of the table, centred across columns, and is considered to be valid in all the columns to the left or right until a different requirement is explicitly reported.}
\label{tab:regions_def}
\centering
 
\begin{tabular}{l|cccccc}
\toprule
Variable & SR & \WmuCR & \WeCR & \ZrcCR & \FakesCR & \LowMetVR \\
\hline
$\pt(j_1)$ [\GeV] & \multicolumn{6}{c}{$>$ 60} \\
$\pt(j_2)$ [\GeV] & \multicolumn{6}{c}{$>$ 50} \\
$\pt(j_{>2})$ [\GeV] & \multicolumn{6}{c}{$>$ 25} \\
\njet &    \multicolumn{6}{c}{2,3} \\
$N_{b\text{-jet}}$ &    \multicolumn{6}{c}{$<$ 2} \\
$|\detajj|$ & \multicolumn{6}{c}{$>$ 3.0} \\
$\eta(j_1)\times\eta(j_2)$ & \multicolumn{6}{c}{$<$ 0} \\
$C_3$ & \multicolumn{6}{c}{$<$ 0.7} \\
$\Delta\phi(j_{i},\metnolepvec)$ & \multicolumn{6}{c}{$>$ 1.0} \\
$N_\gamma$ & \multicolumn{6}{c}{1} \\
$\dphijj$ & \multicolumn{6}{c}{$<$ 2.5 [2.0]} \\
$\Delta\phi(\gamma,\metnolepvec)$   & \multicolumn{6}{c}{$>$ 1.8 [--]} \\
$\pt(\gamma)$ [\GeV] & \multicolumn{6}{c}{$>$ 15, $<$ 110 [$>$ 15, $<$ max(110, 0.733 $\times~\MT{}$)]} \\
\mjj[\TeV] & \multicolumn{5}{c}{$>$ 0.25} & 0.25--1.0\\
\etjetnojvt[\GeV] & \multicolumn{5}{c}{$>$ 130} & $>$ 100\\
\MET[\GeV]  & $>$ 150 & -- & $>$ 80 & $>$ 150 & $<$ 80 & 110--150 \\
\metnolep[\GeV] & -- & $>$ 150 & $>$ 150 & -- & $>$ 150 & 110--150 \\
$C_\gamma$  & $>$ 0.4 & $>$ 0.4 & $>$ 0.4 & $<$ 0.4 & $>$ 0.4 & $>$ 0.4 \\
$N_{\ell}$ & 0  & 1 $\mu$ & 1 $e$ & 0 & 1 $e$  & 0  \\
$\pt(\ell)$ [\GeV]  & --  & $>$ 30 & $>$ 30 & -- & $>$ 30 & -- \\
\bottomrule
\end{tabular}
\end{table}
 
\subsection{Baseline event selection}
\label{subsec:srdef}
 
Stringent background discrimination is made possible by the characteristic features of the VBF process, such as the presence of two highly energetic jets, typically in opposite $\eta$ hemispheres of the detector, more forward than jets from non-VBF processes at comparable momentum transfer of the initial-state partons. These features lead, for example, to large values of the pseudorapidity separation \detajj and invariant mass \mjj of the jet pair. Multijet production predicted by QCD is instead characterized by two back-to-back leading jets in the transverse plane ($\dphijj\sim\pi$); therefore, to reduce that background contribution, the azimuthal separation of the leading jets, $\dphijj$, is required to be smaller than $2.5$ for all SRs.
 
The requirements defining the different regions considered in the analysis are summarized in Table~\ref{tab:regions_def}. The following section describes the selection criteria.
 
The photon produced in association with either the Higgs boson (for the \Hinv search) or the \Zboson boson (for the EW \Zg{} cross-section measurement) is usually radiated from the scattering \Wboson bosons, and it is produced within the large rapidity gap between the two leading jets. For this reason the photon centrality $C_\gamma$~\cite{HIGG-2013-13} is defined as
\begin{equation}
C_\gamma=\text{exp}{\bigg[}-
\frac{4}{(\eta_1-\eta_2)^2}{\bigg(}\eta_\gamma-\frac{\eta_1+\eta_2}{2}{\bigg)}^2{\bigg]},
\label{eq:centrality}
\end{equation}
where the subscripts 1 and 2 indicate the highest- and second-highest \pt jets in the event. The value of $C_\gamma$ is 1 when the photon is centred between the two jets characterizing the VBF signature,$1/e$ when it is aligned with one of the two jets, and tends to zero when the photon is farther forward in $|\eta|$ than the jets.
 
Events are assigned to the SR if they satisfy a set of requirements that have been optimized to maximize the sensitivity of the analysis to the EW \Zg{} and VBF \Hinv{} signals. These requirements are summarized in the following.
 
As discussed in \cref{sec:data}, events are selected with the \met{}-trigger algorithm. To ensure a trigger efficiency exceeding 97\% in this topology, the offline \MET{}, after full offline reconstruction and calibration of all the objects in the events, is required to be larger than 150~\GeV{}. Corrections and uncertainties in those corrections to the simulation modelling of the \met{} trigger are discussed in \cref{subsec:systematics}.
 
The leading and subleading jets are required to have $\pt > 60~\GeV{}$ and 50~\GeV{}, respectively, both satisfying the fJVT requirements mentioned in Section \ref{sec:objects}. These same two leading jets must be located in opposite $\eta$ hemispheres ($\eta(j_1)\times\eta(j_2) <$ 0), to be well separated in pseudorapidity ($|\detajj| > 3.0$) and not back-to-back in the plane transverse to the beamline ($\dphijj < 2.5$), and must have large invariant mass ($\mjj > 0.25~\TeV{}$).
 
Another characteristic of the EW processes in the VBF topology is reduced hadronic activity in the large rapidity gap between the two leading jets, caused by the absence of colour connection between the two quarks. To suppress the contribution from strong \Vgamma production with additional jets from QCD radiation in comparison with the benchmark signals, the equivalent centrality $C_3$ is defined for the third-leading jet in the event, if any, replacing $\eta_\gamma$ in Eq.~\eqref{eq:centrality} with the third-leading jet's pseudorapidity $\eta_3$. A third jet with $\pt > 25~\GeV{}$ can be present in the events, and additional jets are allowed only if they have $\pt < 25~\GeV{}$. The third-leading jet is required to be in one of the two forward regions, corresponding to a small centrality value, $C_3 < 0.7$.
 
All jets must be well separated in azimuth from the \metvec{} direction, satisfying $\Delta\phi(j_{i},\metvec{}) > 1$ for each jet $j_{i}$ with $\pt > 25~\GeV{}$. At most one \bjet identified using the algorithm defined in Section~\ref{sec:objects} is allowed to be present, and the \etjetnojvt variable must be larger than 130 \GeV{} in each event. The requirement on the maximum number of \bjets has a negligible effect (<0.1\%) on the signal selection efficiencies, while it ensures that the dataset considered in the search for \Hinv{} produced through VBF presented in this paper is orthogonal to the dataset considered in the search for invisible decays of Higgs bosons produced in association with \ttbar~\cite{SUSY-2018-12}.
 
Each event must contain a single reconstructed photon with $15 < \pt < 110~\GeV{}$, $\dphimetgam > 1.8$, and $C_\gamma > 0.4$. The upper bound on the photon \pT{} reduces contributions from background processes, especially the \gjet{} background. The $z$ coordinate of the reconstructed photon extrapolated from the calorimeter to the beamline must be no more than 250~mm from the identified primary vertex to reduce the number of photons from non-collision backgrounds.
 
In the SR, all events with leptons are vetoed, specifically ensuring the absence of muon or electron candidates fulfilling the criteria described in Section~\ref{sec:objects}. Electron and muon candidates considered in this veto are not required to satisfy any isolation or impact parameter requirement, and muons are considered before the overlap removal requirement.
 
\subsection{Event classification and fiducial-volume definition for EW \Zg{} measurement}
\label{subsec:mjjbinning}
 
In the signature considered in this analysis, the \mjj{} observable helps in discriminating between the strong and EW \Zg{} processes. The strong production \Zg{} process tends to have a smaller \mjj{} than the EW \Zg{} process.
In the context of the EW \Zg{} cross-section measurement, the events satisfying the baseline SR requirements described in Section~\ref{subsec:srdef} are therefore binned according to their \mjj{} value, with bin boundaries of 250~\GeV{}, 500~\GeV{}, 1000~\GeV{} and 1500~\GeV{}. The highest \mjj{} in the selected data events is 4.0~\TeV{}.
 
The fiducial volume for the EW \Zg{} cross-section measurement is defined to closely follow the SR definition reported in Section~\ref{subsec:srdef}. Particle-level requirements are placed on stable particles (defined as having a mean lifetime $c\tau > 10$~mm) before interaction with the detector but after hadronization. The four-momenta of prompt leptons, i.e. those which do not originate from the decay of a hadron, include the sum of the four-momenta of prompt photons within a cone of size $\Delta R = 0.1$ around the lepton, and the resulting leptons are referred to as `dressed' leptons. Jets are constructed using an anti-$k_t$ algorithm with radius parameter of 0.4, excluding electrons, muons, neutrinos, and photons associated with the decay of a $\Wboson$ or $\Zboson$ boson. Photon isolation momentum $E_{\text{T}}^{\text{cone20}}$ is defined as the transverse momentum of the system of stable particles within a cone of size $\Delta R = 0.2$ around the photon, excluding muons, neutrinos, and the photon itself. The photon isolation momentum is required to be less than 7\% of the photon transverse momentum. The truth-\metvec{} is defined as the vector sum of the transverse momenta of all neutrinos in the simulated final state, including those resulting from hadronization.

A veto on dressed muons or electrons with $\pT > 4~\GeV{}$ and $|\eta| < 2.47$ is applied. Jets and photons are required to be separated in angular distance, with $\Delta R(j, \ell) > 0.4$.
The remaining requirements, similar to the SR ones, are summarized in Table~\ref{table:lep_sel_fid}, and the acceptance times reconstruction efficiency for the fiducial-volume definition is 0.4\%.
 
\begin{table}[!htbp]
\caption{Fiducial-volume definition for the EW \Zg{} cross-section measurement.}
\label{table:lep_sel_fid}
\centering
\begin{tabular}{l|c}
\toprule
Observable & Requirements\\
\hline
\njet with $\pT > 25~\GeV{}$ & $\geq 2$ \\
$|\eta(j_{1,2})|$ & $<$ 4.5 \\
$\pt(j_1)$ [\GeV] & $>$ 60 \\
$\pt(j_2)$ [\GeV] & $>$ 50 \\
$\Delta R(j, \ell)$ & $>$ 0.4 \\
$|\detajj|$ & $>$ 3.0 \\
$C_3$ & $<$ 0.7 \\
\mjj{} [\TeV] & $>$ 0.5 \\
truth-\MET{} [\GeV] & $>$ 150\\
$\Delta\phi(\text{truth-}\metvec, j_i)$ & $>$ 1.0\\
$\pt(\gamma)$ [\GeV] & $>$ 15, $<$ 110 \\
$|\eta(\gamma)|$ & $<$ 2.37 \\
$E_{\text{T}}^{\text{cone20}}/E_{\text{T}}^\gamma$ & $<$ 0.07 \\
$\Delta R(\gamma, \text{jet-or-}\ell)$ & $>$ 0.4 \\
$C_\gamma$ & $>$ 0.4 \\
$\Delta\phi(\text{truth-}\metvec, \gamma)$ & $>$ 1.8\\
$N_\ell$ with $\pT > 4~\GeV{}$ and $|\eta| < 2.47$ & 0 \\
\bottomrule
\end{tabular}
\end{table}
 
\subsection{Event classification for \Hinv{} search}
\label{subsec:keras}
 
In the context of the search for \Hinv{} signal, a Dense Neural Network (DNN), based on the \keras library~\cite{chollet2015keras,franoischollet2017learning} with \textsc{Tensorflow}~\cite{abadi2016tensorflow} as a backend, was designed and trained to separate such signal from the background contributions by using kinematic features. The DNN assigns signal-like (background-like) events an output score close to 1 (0).
 
The DNN architecture is composed of three blocks, each of which includes a Dense layer with 384 neurons, a Dropout rate and a Batch Normalization.
The Rectified Linear Unit (ReLU) activation function~\cite{10.5555/3104322.3104425} is used for each Dense layer, while a sigmoid activation function is used for the output node. To regularize the training and avoid overtraining, the Dropout and Batch Normalization layers are included in the DNN architecture.
Each of the three blocks receives as input the output of the previous block concatenated with the input features.
 
The DNN is trained using a mixed sample of signal and background events, weighted according to their expected yields in the considered final state.
To increase the number of events available for the training, the selection criteria defining the training sample phase space are looser than those for the SR and CRs. Events used for the training are required to have either two or three jets satisfying the identification criteria described in Section~\ref{sec:objects}, a single photon, and $\metnolep > 140~\GeV{}$.
The sample for each considered process is split into a training sample and a testing sample composed of 85\% and 15\% of the original sample, respectively. The training sample is further divided into a subsample used in the proper training (80\%) and a sub-sample for validation (20\%).
 
During the DNN training the weights are updated via an optimizer used to minimize the binary Cross-Entropy loss, while at each step a customized metric is evaluated, representative of the expected limit $\mathcal{E}$ on the \Hinv{} signal. This expected-limit approximation is defined below and the expected-background error in the numerator includes the background yield, the MC statistical uncertainties, and the data background uncertainty in data as provided by the CR:
\begin{equation*}
\label{eq:ele}
\mathcal{E}^{-1} = \sqrt{\sum_{i\in \text{bins}} \mathcal{E}_i^{-2}}~, \qquad \mathcal{E}_i = 2\frac{\sqrt{b_i + \left( b_i^{V\gamma}\cdot\frac{\sqrt{d_i(\mathrm{CR})+\sum_{j\in \mathrm{ev}} w_{i,j}(\mathrm{CR})}}{b_i^{V\gamma}(\mathrm{CR})} \right)^2 + \sum_{j\in \mathrm{ev}} w_{i,j}^2}}{s_i}~\mathrm{,}
\end{equation*}
where $b_i$ is the total background yield in bin $i$ of the DNN output distribution, $s_i$ is the signal yield, $b_i^{V\gamma}$ is the contribution to the background from \Wg{} and \Zg{} processes, $d_i(\mathrm{CR})$ is the number of events in data in the one-lepton CR defined in Section~\ref{subsec:crdef}, and $w_{i,j}$ is the MC weight of each event ($\mathrm{ev}$) normalized to the integrated luminosity summed over the events in the considered bin.
 
After each training epoch the DNN weights are stored only if $\mathcal{E}$ is improved and the difference in loss between the training and validation samples is smaller than 3\%, which prevents failures from overtraining.
 
After the composition of the training samples and the DNN architecture were established, a backward elimination procedure was carried out to reduce the number of input features with minimal performance loss. Input features are varied up and down by 10\% of their nominal values while evaluating the impact of each as the change in the mean DNN output score. The most relevant input features are expected to have a greater impact on the output if their value is modified, while the least relevant ones will produce smaller variations, and hence can be dropped with no significant loss. When an input feature is dropped the DNN is retrained with the new subset of input features, and the procedure is repeated while the relative change in $\mathcal{E}$ from its value with that with all input features is less than 2\%. The remaining features have a non-negligible impact on the DNN; hence they are used in the analysis.
The 8 most significant kinematic features were selected: the separation of two leading jets in $\eta$, \detajj, and in $\phi$, \dphijj; the dijet invariant mass \mjj; the photon pseudorapidity $\eta(\gamma)$; the two leading jet transverse momenta $p_{T}^{j_1}$ and $p_{T}^{j_2}$; the subleading jet pseudorapidity $\eta(j_2)$; and the lepton-subtracted missing transverse momentum \metnolep{}. When their nominal value is perturbed by 10\%, \detajj{}, \metnolep{} and \mjj{} produce the largest change in DNN output score; therefore, they are interpreted as the most important in separating signal from background events.
 
All events are categorized according to DNN output score into four bins chosen to optimize $\mathcal{E}$, subject to the requirements that each bin be at least 0.05 units wide in output score and that there be at least 20 background events expected in each bin in order to have sufficient background yield in the CRs to validate the predictions. The optimal boundaries of the DNN output score bins are $[0.0,0.25,0.6,0.8,1.0]$. The acceptance times reconstruction efficiency for VBF Higgs boson production with an additional photon is around 0.7\%, which comes from the high \met{} threshold as well as the other selections.
 
\subsection{Event classification for \Hyyd{} search}
\label{subsec:mtbinning}
 
The \Hyyd decay presents a striking signature characterized by a kinematic edge at the Higgs boson mass in the distributions of the transverse mass constructed from the photon and \met, defined as:
\begin{equation*}
\mtgammet = \sqrt{2\pt^{\gamma}\met \left[1-\cos(\phi_\gamma-\phi_{\met})\right]}\ .
\label{eq:MT}
\end{equation*}
 
To increase the search's sensitivity to this semi-visible signal, a dedicated signal region SR$_{\gammad}$ is defined by changing some of the selection requirements in Section~\ref{subsec:srdef}. No requirement on \dphimetgam is applied when targeting this specific benchmark and, for $\mtgammet > 150~\GeV{}$, the upper bound on the photon \pt is relaxed to $0.733\times\mtgammet$. The requirement on $\dphijj$ is tightened, so that all events have $\dphijj < 2.0$.
These selections are changed for both the SRs and the CRs defined later in Section~\ref{subsec:crdef}.
The events satisfying the SR$_{\gammad}$ requirements are therefore binned according to \mtgammet{} value, with bin boundaries of 0~\GeV{}, 90~\GeV{}, 130~\GeV{}, 200~\GeV{} and 350~\GeV{}.
The events are further separated according to the value of the dijet invariant mass \mjj, into events with $\mjj < 1.0~\TeV{}$  and events with $\mjj{} \geq 1.0~\TeV{}$. These two \mjj regions have different relative contributions of the VBF- and ggF-produced Higgs boson signals, so the separation improves the overall sensitivity to the \Hyyd decay. The acceptance times reconstruction efficiency for VBF \Hyyd{} production is 0.4\% for a 125~\GeV{} scalar mediator but increases to 3.1\% for a 500~\GeV{} scalar mediator.
 
\subsection{Control region definitions}
\label{subsec:crdef}
 
Mutually exclusive CRs are defined in order to constrain the normalization and validate the modelling of kinematic distributions for the two dominant backgrounds from \Wgamma and strong \Zgamma production. These CRs have kinematic features similar to those in the SR but are dominated by the background processes.
The EW production of \Zgammajj events, which is measured in this paper and constitutes an irreducible background contribution in the search for invisible and semi-visible decays of the Higgs boson, is mitigated through the event classification applied to SR events described in Sections~\ref{subsec:keras} and \ref{subsec:mtbinning}.
 
The \WmuCR and \WeCR are defined by changing the zero-lepton requirement for the SR to instead require exactly one lepton, either a muon or an electron, respectively. Data events in the region characterized by one electron are collected using a single-electron trigger algorithm, while those characterized by one muon are collected using either the same \met trigger algorithm used to select SR events or a single-muon trigger algorithm. In both cases the leptons are required to have $\pt > 30~\GeV{}$ in order to be on the single-lepton trigger efficiency plateau.
 
In each of these CRs the presence of a single muon (electron) satisfying the medium (tight) identification criteria, passing the loose isolation requirement, and surviving the overlap removal procedure is required. The efficiency of selecting muons (electrons) averaged over $\pt > 30~\GeV{}$ and $|\eta| < 2.7$ (2.47) is larger than 96\% (80\%). To ensure that electron and muon candidates originate from the primary vertex, the significance of the track’s transverse impact parameter relative to the beam line must satisfy $|\dzero/\sigma(\dzero)| < 3$ (5) for muons (electrons), and the longitudinal impact parameter $z_0$ is required to satisfy $|\zzsth| < 0.5$~mm.
As done in the SR, the looser identification criteria are applied to define the veto on any additional lepton.

To reduce the contribution from jets faking electrons, $\met > 80~\GeV{}$ is required.
The \FakesCR{}, containing events with $\met < 80~\GeV{}$, is used to estimate the jets-faking-electrons background.
The requirements on the reconstructed jets and photon are the same as for the SR, while requirements on the \met, other than the one specifically for the \FakesCR{} mentioned above, are replaced by equivalent ones on the \metnolep.
 
To check the modelling of the strong \Zgamma background contribution, a \ZrcCR is defined by applying all the requirements of the SR definition but reversing the photon centrality requirement, selecting events with $C_\gamma < 0.4$. There is a 7\% EW \Zgammajj{} contribution to the \ZrcCR, which is taken into account in the statistical analysis of the cross-section measurement for this process.
 
\subsection{Validation region for  \gjet background}
\label{subsec:vrdef}
 
In order to correct the normalization of the \gjet background process and validate the modelling of its kinematic features, an orthogonal sample of events characterized by a small amount of \met in the final state is considered. In this region (\LowMetVR) the same selections as the SR are applied, except that the \met has to be in the range $110 < \met < 150~\GeV{}$, the upper bound on the photon \pt is removed, no \etjetnojvt requirement is applied, and the leading dijet system's invariant mass has to satisfy $0.25 < \mjj < 1.0~\TeV{}$.
 
For the \Hyyd{} search, a similar Low-\met validation region is defined (Low \met-\gammad VR) with the same selection as the \LowMetVR but also requiring $\mtgammet < 110~\GeV{}$ without any selection on \dphimetgam.


\section{Data analysis}
\label{sec:analysis}

The different analyses presented in this paper focus on processes sharing a similar signature. 
For the EW \Zg{} cross-section measurement, all the SM processes contributing to the considered final-state signature are estimated using both the simulated samples, described in Section~\ref{sec:simulation}, and data-driven techniques, described in Section~\ref{subsec:background}.
The \mjj distribution is used to categorize events into SR selections with different signal-to-background ratios, as detailed in Section~\ref{subsec:mjjbinning}.
In the search for invisible Higgs boson decays, the same strategy is adopted to estimate the SM backgrounds but a different observable is considered. A multivariate discriminant is designed to categorize events falling in the SR according to an increasing expected significance for the invisible Higgs boson decay signal contribution relative to the SM background expectation, as detailed in Section~\ref{subsec:keras}.
For the \Hyyd search, events are instead classified according to their kinematic features, as detailed in Section~\ref{subsec:mtbinning}.
Control regions enriched in background contributions, described in \cref{subsec:crdef}, are considered in the statistical analysis of the data, while validation regions are considered only to qualitatively assess the modelling of specific background contributions, as detailed in \cref{subsec:vrdef}.
The theoretical and experimental uncertainties of the signal and background predictions are described in Section~\ref{subsec:systematics}.
 
\subsection{Background contribution estimation}
\label{subsec:background}
 
The dominant contributions entering the analysis SR are \Zgamma and \Wgamma events in which the lepton from the \Wboson decay is lost mostly because it falls outside of the \pt or $\eta$ acceptance. Roughly 28\% of \Wg{} events in the 0-lepton SR come from hadronic $\tau$-lepton decays with truth $\pT>20$~\GeV{} and $|\eta|<2.5$; this fraction is too small to benefit from a veto on reconstructed hadronic $\tau$-lepton decays, which would incur additional uncertainties on their reconstruction efficiency. The uncertainty in the modelling of the lost lepton is included in the MC simulation, as detailed in Section~\ref{subsec:systematics}. In order to validate the theoretical and experimental uncertainties, these major backgrounds are tested in the CRs described in Section~\ref{sec:event_selection} characterized by the same requirements as those of the SR but requiring events with exactly one lepton or reversing the $C_\gamma$ requirement. The CRs with one selected lepton (e.g.\ \WeCR{} and \WmuCR{}) are expected to be dominated by \Wgamma events while the \ZrcCR with the reversed $C_\gamma$ requirement is expected to be dominated by strong production of \Zgamma events. The background yield compared with the data in the aforementioned CRs is utilized to determine the normalization of the \Wgamma and \Zgamma backgrounds by allowing their normalization to float in the different fit models, as described in Section~\ref{sec:fit_models}.
 
An additional background contribution results from $Z\text{\,+\,jets}$ or $W\text{\,+\,jets}$ events in which one of the jets is misreconstructed as a photon. The simulation described in Section~\ref{sec:simulation} is not expected to properly reproduce the rate of jets misreconstructed as photons. Therefore, the $\text{jet}\to\gamma$ misreconstruction rate is estimated using a technique tuned on data amd similar to the one described in Ref.~\cite{STDM-2010-08}.
The method, often referred to as an `ABCD' method, relies on the evaluation of the photon candidate event yield in a two-dimensional plane defined by the amount of transverse energy deposited in clusters of cells within a cone of size $\Delta R = 0.4$ around the photon, excluding the photon cluster itself (i.e. isolation), and by the quality of the photon identification criteria (tightness)~\cite{EGAM-2018-01}. A photon signal region (region $A$) is defined by photon candidates that are isolated and satisfy the tight identification requirements described in \cref{sec:objects}. Three background regions are defined in the isolation--tightness plane, consisting of photon candidates which are tight but non-isolated (region $B$), non-tight but isolated (region $C$) or neither tight nor isolated (region $D$).
A photon candidate is defined as non-isolated if it does not satisfy the requirement on the amount of isolation transverse energy, while it is classified as non-tight if it fails the tight identification but satisfies a modified set of requirements related to four of the selections associated with the shower-shape variables computed from the energy deposits in the first layer of the EM calorimeter.
No correlation between the photon identification and isolation requirements would imply that the numbers of photon candidates in the four regions ($A$, $B$, $C$, $D$) satisfy the condition $N^A/N^B = N^C/N^D$.
Although this condition is almost fully satisfied, the residual correlation between the photon isolation and tightness is taken into account through correction factors estimated from MC simulations, differing from unity by up to 40\%.
A further correction to the described method is included to take into account the contamination from real photons in the three regions which are supposed to be dominated by fake photons ($B$, $C$, $D$). This contribution is evaluated using MC simulation and is parameterized through coefficients representing the fraction of real photons in each of the aforementioned regions relative to the fraction of real photons observed in region $A$.
With these two corrections, this method determines the $\text{jet}\to\gamma$ misreconstruction rate and consequently the contribution due to this background in all the kinematic regions considered in this analysis. The non-tight photon and isolation definitions were varied to compute a relative systematic uncertainty, estimated to be 80--90\% for this background process.
 
The background contribution due to electrons misidentified as photons is estimated through an $e^\pm\to\gamma$ misreconstruction rate (fake rate), determined from a comparison of the rates of \Zboson boson reconstruction in the $e^\pm\gamma$ and $e^+e^-$ final states, as in Ref.~\cite{EXOT-2016-32}.
This background is small in the EW \Zg{} measurement and in the Higgs boson invisible decays interpretation but is more relevant in the dark-photon interpretation.
The full Run 2 dataset is used to select \Zee events in which the electron pairs in the final state are either reconstructed as an $e^+e^-$ pair or misreconstructed as an $e^\pm\gamma$ pair. The invariant mass \mee or \meg is required to be consistent with the \Zboson boson mass within 10~\GeV. The yields of $Z$ events are then obtained from a simultaneous fit to the \mee or \meg distributions, in order to subtract the contamination from jets misidentified as electrons or photons in the two samples using an extrapolation from the sidebands of the pair's mass. The $e^\pm\to\gamma$ misidentification rate, measured as a function of \abseta and \pT, varies between 1.5\% and 9\%, with larger values associated with larger values of \abseta, since the misidentification rate depends on the amount of material in front of the calorimeter. The background contribution from misidentified electrons is evaluated in the different regions considered in this analysis by applying the calculated $e^\pm\to\gamma$ misidentification rates to event yields resulting from the same criteria except that an electron is required instead of a photon. The misidentification rate includes uncertainties from varying the fitted range of the pair mass, turning the jet background subtraction on and off, and accounting for biases in the energy of the photon compared to that of the electron. The total uncertainties vary from 30\% for $\pt = 15~\GeV{}$ to 15\% for $\pt = 100~\GeV{}$.
 
The \gj background is a minor one in this analysis since these events do not have intrinsic \met, other than the small contribution from the presence of neutrinos in heavy-flavour jet decays. While only a small fraction of such events exceed the \met requirement of the SR definition, most of this background contribution results from jet mismeasurement, yielding a significant amount of misreconstructed \met.
The sample of simulated events for this large cross-section process is limited, so a jet smearing approach is used to increase the sample size by a factor of 20.
A simulated sample of \gj events is filtered with a set of criteria looser than those used in the SR definition so that events include one identified photon and two or more reconstructed jets with $|\detajj|>2.5$. The leading and subleading reconstructed jet \pt must be larger than 50 and 30~\GeV, respectively.
The energy of each truth jet, which includes neutrino momenta, in these events is smeared according to the jet energy response evaluated in bins of \pt of the corresponding truth-level jets; jet $\phi$ and $\eta$ are smeared with a Gaussian distribution according to the angular resolution of PFlow jets as measured in 13~\TeV{} data~\cite{ATLAS-CONF-2018-038}. The \pileup jet energies are not smeared, but they are included in the jet counting.
After all the truth jets have been smeared, the corresponding JVT and fJVT tagging scores are recalculated for each jet. The true-\met magnitude and direction due to heavy-flavour decay neutrinos are resampled from simulated sample truth information. In addition to this redefinition of the \met in the event, corrections due to jet smearing are also propagated to the reconstructed \met as well as all other kinematic quantities used to define the SR.
The \gj sample obtained with this procedure is normalized according to the simulation-to-data comparison in the \LowMetVR described in \cref{sec:event_selection}, after subtracting all the other background contributions in that region. Normalization to data absorbs any residual contributions of detector noise or other misreconstructed jets. The resulting normalization, which was found to be statistically consistent over all data-taking years, is $0.91\pm0.36$ with the uncertainty coming from the statistical uncertainties of the data. Uncertainties in the extrapolation from the \LowMetVR{} are computed using the difference between the \met{} trigger efficiencies for \gj{} and \Zgamma{} events, which results in a 75\% uncertainty in yield. Lastly, the smearing parameterizations are varied within their uncertainties, and the truth-\MET{} resampling is turned on and off. The total uncertainty in this background is roughly 90\%.
 
Some of the events in the \Wgamma CR are events where a jet or a photon is misidentified as a lepton. Events of this kind result from \gj{} production in which a jet is misidentified as a lepton, diphoton production where a photon mimics the prompt electron, or to multijet events where one jet mimics the photon and another one mimics the lepton. The contribution from fake-lepton events is found to be negligible in the case of events with one muon. To evaluate this contribution to the \WeCR, a corresponding CR (\FakesCR) is defined, as discussed in \cref{subsec:crdef}. The ratio of the yields of fake electrons in the \WeCR and \FakesCR, labelled $\mathcal{R}_{\textrm{fake-}e}$, is estimated in the same regions except that the one electron satisfying the tight identification requirements is replaced by an electron satisfying a loose identification requirement but failing the tight \WeCR{} electron definition. This is called the anti-ID selection. The number of data events in the anti-ID \WeCR, after subtracting residual contributions from prompt-lepton events using simulation, is divided by the number of data events in the anti-ID \FakesCR{} to compute $\mathcal{R}_{\textrm{fake-}e}$. The ratio $\mathcal{R}_{\textrm{fake-}e}$ is measured to be $0.14 \pm 0.11$ in the baseline selection (\cref{subsec:srdef}) and $0.26 \pm 0.10$ in the selection used in the \Hyyd{} search (\cref{subsec:mtbinning}), with the uncertainties coming from statistical uncertainties of the data. This ratio is used to scale the fake-electron contributions in the \FakesCR to obtain the expected background in the \WeCR{}.
 
\subsection{Systematic uncertainties}
\label{subsec:systematics}

Theoretical uncertainties affect all simulated signal and background processes. They originate from the limited order at which the matrix elements are calculated, the matching of those calculations to parton showers, and the uncertainty of the proton PDFs.
 
For the \Zg{} and \Wg{} processes the higher-order matrix element effects and parton-shower-matching uncertainties are assessed by varying the renormalization and factorization scale choices used in the event generation. The effect of resummation and CKKW matching scale~\cite{Lonnblad:2011xx} choices are found to be negligible.
The factorization and renormalization scales are varied up and down by a factor of two using event weights varied in the \SHERPA MC simulation of strong production, while they are varied in separate calculations in \VBFNLO{} for EW production. The corresponding uncertainties are calculated by taking an envelope of the seven factorization/renormalization scale variations: the central value, each scale independently varied up/down, and both scales coherently varied up/down.
For the strong-production \Vg{} background, the effect of the 7-point scale variations on the expected yield ranges from $\sim$25\% to $\sim$56\% in the kinematic regions considered; the corresponding values for the EW \Vg{} process range from $\sim$3\% to $\sim$11\%. Since the EW \Vg{} simulation is leading-order \MGNLO reweighted by \NLO \VBFNLO, the 7-point scale variations are computed from \VBFNLO{}.
For the strong-production \Vg{} background the scale variations are propagated to both the matrix element and the parton shower, while for the EW \Vg{} samples the \PYTHIA[8] A14 `eigentune' variations~\cite{ATL-PHYS-PUB-2014-021} are summed in quadrature to assess the uncertainty in the parton shower modelling. The systematic uncertainty is found to be in the range 4--15\%.
 
For the \Zgamma{} process, the interference between EW and strong production is computed at LO in $\MGNLO\,+\,\PYTHIA[8]$, as discussed in \cref{subsec:Vggeneration}. The full prediction of the interference sample is assigned to the EW \Zgammajj{} sample as a one-sided uncertainty, as is done in Ref.~\cite{STDM-2017-26}, in all SR and CR bins, and its largest contribution is at large \mtgammet{} with a $-22$\% uncertainty in the prediction without interference.

An uncertainty due to the choice of MC generator is evaluated by comparing strong \Vg{} samples produced with \SHERPA and \MGNLO: the full difference between \SHERPA and \MGNLO is considered as a systematic uncertainty, and the difference in the predicted yields is symmetrized around the \SHERPA-predicted central value for this background yield. The difference is as large as 20\% depending on the signal region or control region bin, but the biggest differences are at large \mtgammet{}. Both the shape and normalization differences are accounted for in the fit model.
 
The PDF uncertainties of the \Wg{} and \Zg{} backgrounds are evaluated in each region and each bin of the SR as the standard deviation of the 100 PDF replicas of the \nnpdf set. The overall uncertainty is found to be smaller than a few percent.
 
For the $\text{Higgs}\to\gamma\gammad$ signal samples, the VBF and ggF Higgs boson inclusive production cross-sections and uncertainties are provided by the LHC Higgs Cross Section Working Group~\cite{deFlorian:2016spz}.
For the VBF process, \mjj{}-dependent renormalization and factorization uncertainties and their correlations were computed by the LHC Higgs working group~\cite{deFlorian:2016spz} for the \POWPY[8] signal samples. The parton shower uncertainty for the VBF signal is estimated to range from 2\% to 4\% by comparison with a \POWHER[7] sample. There is a 2\% uncertainty due to the Higgs boson \pT{}-dependent \NLO electroweak correction, which is estimated from~\hawk~\cite{Denner:2014cla}. Uncertainties due to the PDF choice are assessed from the average variations of the \textsc{PDF4LHC15} set of PDFs.
For the ggF process, uncertainties from the renormalization and factorization scale variations are also assigned in categories of Higgs boson \pT{}, number of jets, and \mjj{} by the LHC Higgs Cross Section Working Group~\cite{deFlorian:2016spz}. The resulting uncertainties are 20\%--30\%. The smaller PDF and parton shower variations are also included. The same uncertainty treatment is used for the small contribution of ggF-produced \Hinv{} events.
 
For the VBF Higgs$+\gamma$ sample, the factorization and renormalization scales are varied up and down by a factor of two through event weights varied in the MC matrix element and parton shower simultaneously. The envelope of the seven factorization/renormalization scale variations considered is assigned as the corresponding 1--2\% systematic uncertainty. The uncertainty due to the modelling of the parton showering is assessed by comparing relevant kinematic distributions for \PYTHIA[8] and \HERWIG[7] samples generated at LO in \alphas{}, as discussed in \cref{sec:sig_process}. The relative differences of 2\%--8\% are applied to the NLO signal sample as uncertainties. Uncertainties of 1\%--2\% due to the PDF choice are assessed from the average variations of the \textsc{PDF4LHC15} set of PDFs.

Several experimental uncertainties impact the sensitivity of the analyses presented in this paper. They are grouped into categories: uncertainties in the luminosity, uncertainties in the trigger efficiencies, and uncertainties related to the reconstruction of physics objects such as electrons, muons, jets, and the \met. A summary of the systematic impact on the measured signal strength or limits set by this search is reported in Table~\ref{tab:LimitKerasImpact}.
 
The uncertainty in the luminosity is 1.7\%~\cite{ATLAS-CONF-2019-021} and impacts the signal yield and the simulated background yield.
 
The \met requirement of 150 \GeV{} in the SR is key to maximizing the sensitivity of the analysis. The \met triggers are not fully efficient, so systematic uncertainties are used to account for possible trigger efficiency differences between data and simulation. This is done by comparing the combined L1+HLT trigger efficiency as a function of \metnolep{} for simulated \Wmn{} events and data events, both selected with single-muon triggers. Neither the \met trigger nor the offline \met{} reconstruction includes muon momenta in their calculations. The \MET{} trigger has an efficiency of roughly 81\% at $\met{} = 150~\GeV{}$ and ${>}99$\% for $\met > 200~\GeV{}$ in simulation. The \met{}-trigger efficiencies for simulated \Wmn{}, \Wmugamma, and \Zgamma are found to be statistically consistent in all the considered regions, so the same correction for differences between the data and simulation trigger efficiencies is applied to all simulated events passing the \MET{} trigger. The correction as a function of \MET{} varies from around (3--6)$\% \pm 4\%$ at $\met{} = 150~\GeV{}$ to less than $0.4\% \pm 1\%$ for $\met > 200~\GeV{}$ depending on the specific \MET{}-trigger algorithm used in a given data-taking period. The uncertainty in the correction comes from the data statistical uncertainties used to derive the correction. For the \WCR scale factors and uncertainties in events passing lepton triggers, the corresponding single-lepton triggers corrections are applied~\cite{TRIG-2018-01,TRIG-2018-05}.
 
Systematic uncertainties are calculated for lepton reconstruction and isolation efficiencies~\cite{EGAM-2018-01,MUON-2018-03} and for the energy scale and resolution~\cite{EGAM-2018-01}.
For the electron (muon) veto, an uncertainty in the electron (muon) reconstruction inefficiency is taken into account.
For jets, uncertainties are derived in the energy scale and resolution~\cite{JETM-2018-05} and for the \pileup tagging efficiencies~\cite{ATLAS-CONF-2014-018,TRIG-2018-01}.
For the photon, uncertainties in the reconstruction, isolation, energy scale and resolution~\cite{PERF-2017-02,EGAM-2018-01} are considered.
The above uncertainties associated with the reconstructed objects are propagated to the calculation of \met, as is the uncertainty in the scale and resolution of the \met soft term~\cite{ATLAS-CONF-2018-023}.


\section{Fit models and results}
\label{sec:fit_models}

The statistical analysis carried out has two objectives: first measuring the EW \Zg{} production cross-section, and then searching for evidence of BSM physics in specific models involving invisible or semi-visible decays of the Higgs boson.
The different processes of interest are compared to data using a profile-likelihood-ratio test statistic in a frequentist approach.
A maximum-likelihood fit to the observed data in each bin of the event classification described in Sections \ref{subsec:mjjbinning}, \ref{subsec:keras}, and \ref{subsec:mtbinning} is used to set constraints on the signal strength $\mu$ for each model considered, all using asymptotic formulae~\cite{Cowan:2010js}: a two-sided confidence level (CL) interval with the CL$_\text{s}$ definition~\cite{Read:2002hq} is extracted for the EW \Zg{} cross-section measurement, while one-sided confidence levels calculated with the same approach are considered to set upper limits on new physics contributions.
 
Each bin $i$ of the SR is assumed to include a number of events $N_i$ where the total expected background yield is $B_i$ and the signal contribution is $S_i$. A likelihood function $\mathcal{L}$ is defined as
\begin{equation}
\mathcal{L} = \prod_{i\in \mathrm{bins}}\mathcal{P}(N_i|B_i(\overrightarrow{\theta}) + \mu S_i(\overrightarrow{\theta}))\cdot \prod_{j\in\mathrm{syst.}}\mathcal{G}(0|\theta_j)
\qquad
B_i(\overrightarrow{\theta}) = \sum_{k\in \mathrm{bkg. comp.}} \beta_{i,k}\cdot B_{i,k}(\overrightarrow{\theta})
\label{eq:likelihood}
\end{equation}
where $\mathcal{P}(x|\nu)$ is the Poisson probability density function, $\mathcal{G}(x|\theta)$ is the probability density function of a Gaussian with unit width, and $\theta_j$ represents the nuisance parameters corresponding to each considered uncertainty.
The expected background yield in a given bin, $B_i$, is given by the sum of several contributions, where a normalization factor $\beta$ for each background component is included.
The $\beta$ factors represent the overall normalization of a given background contribution estimate.
In the maximum-likelihood fit such normalization factors can be either fixed to 1 or left floating in the maximization. The treatment of these factors in the statistical analysis framework is detailed in the following for each scenario considered.
In the framework of the EW \Zg{} cross-section measurement, this process is considered as the signal and its corresponding $\beta$ factor is replaced by the parameter of interest \muZgEWK{}. In the framework of the searches for invisible or semi-visible decays of the Higgs boson, expected signal yields are evaluated with the assumption $\Binv=1$ or $\Byyd=1$, which allows the parameter of interest $\mu$ to directly determine the actual Higgs boson decay branching fraction to the considered invisible or semi-visible particles.\footnote{When small signal contaminations are expected in a CR, these are correctly accounted for in the statistical analysis.} The signal considered in this framework is normalized to the SM cross-section for Higgs boson production.
The expected yields depend not only on $B_i$ and $S_i$ but also on the nuisance parameters, although these parameters are constrained in the likelihood function by the $\mathcal{G}(0|\theta_j)$ factors.
Some of the systematic uncertainties affecting background and signal predictions vary the final observable by less than 0.1\% and these are ignored to improve the stability of the fit with no loss of accuracy.
 
Each experimental uncertainty source is taken to be fully correlated across all signal and control regions. The applied correlation of theory systematic uncertainties depends on the type of uncertainty. PDF uncertainties are treated as fully correlated across bins. The uncertainty in the interference between the EW and strong \Zgamma{} processes is treated as a fully correlated one-sided uncertainty in the EW \Zgammajj{} process. The parton showering in each of the four \Vg{} samples (EW and strong-production components of \Wg{} and \Zg{}) is correlated across all bins and similar for the scale variations in the strong-production \Vg{} samples, except the scale uncertainties in the one-lepton CRs which are not correlated with the other bins. A separate nuisance parameter for the latter is motivated by the less than 1\% background contribution to the one-lepton CR, and the contribution due to missing one of the leptons. Scale uncertainties for EW \Zgjj{} and EW \Wgjj{} are treated as uncorrelated among regions and correlated in bins of the same region to avoid constraining this source of uncertainty. Within rounding, no change in the measured significance nor in the extracted limits was found compared with correlating these uncertainties across all analysis bins.
 
\subsection{Fit model and results for the EW \Zg{} cross-section measurement}
\label{subsec:zgewkfitmodel}
The signature considered in this paper has a significant contribution from EW \Zgammajj{} production, which has not been observed previously. A measurement of the cross-section for this process is an important first step, as a result complementary to the prior SM CMS observation of EW \Zlljj{}~\cite{cmscollaboration2021measurement}. No BSM signal contributions are considered in this subsection.
 
For this measurement events are categorized into four \mjj bins, as described in Section~\ref{subsec:mjjbinning}, to profit from differences between the kinematic distributions of EW \Zg{} production and strong \Zg{} production. The statistical analysis of data entering either the SR, \WmuCR, \WeCR, \ZrcCR, or \FakesCR is performed according to their \mjj bins, expanding the likelihood function defined in \cref{eq:likelihood} as
\begin{align}
& \mathcal{L}(\muZgEWK,\beta_{\Zboson\gamma_{\mathrm{strong}}},\beta_{\Wboson\gamma},\overrightarrow{\theta}) = \nonumber \\
& \prod_{i\in \mathrm{SR~bins}}\mathcal{P}(N_i|\muZgEWK S_{i,\Zboson\gamma_{\mathrm{EW}}} +
\beta_{\Zboson\gamma_{\mathrm{strong}}} B_{i,\Zboson\gamma_{\mathrm{strong}}} +
\beta_{\Wboson\gamma} B_{i,\Wboson\gamma} + B_{i,\mathrm{others}})  \nonumber \\
\times~  & \prod_{k\in \mathrm{CR~bins}}\mathcal{P}(N_k^{\WeCR}|\beta_{\Wboson\gamma} B_{k,\Wboson\gamma}^{\WeCR} + B_{k,\mathrm{others}}^{\WeCR} +
\beta_{k,\textrm{fake-}e} B_{k,\mathrm{fake-}e}^{\FakesCR} \mathcal{R}_{\textrm{fake-}e}) \nonumber\\
\times~  & \prod_{k\in \mathrm{CR~bins}}\mathcal{P}(N_k^{\WmuCR}|\beta_{\Wboson\gamma} B_{k,\Wboson\gamma}^{\WmuCR} + B_{k,\mathrm{others}}^{\WmuCR}) \nonumber\\
\times~  & \prod_{k\in \mathrm{CR~bins}}\mathcal{P}(N_k^{\ZrcCR}|\muZgEWK S_{k,\Zboson\gamma_{\mathrm{EW}}}^{\ZrcCR} +
\beta_{\Zboson\gamma_{\mathrm{strong}}} B_{k,\Zboson\gamma_{\mathrm{strong}}}^{\ZrcCR} +
\beta_{\Wboson\gamma} B_{k,\Wboson\gamma}^{\ZrcCR} +
B_{k,\mathrm{others}}^{\ZrcCR})  \nonumber\\
\times~  & \prod_{k\in \mathrm{CR~bins}}\mathcal{P}(N_k^{\FakesCR}|\beta_{\Wboson\gamma} B_{k,\Wboson\gamma}^{\FakesCR} + B_{k,\mathrm{others}}^{\FakesCR} + \beta_{k,\textrm{fake-}e} B_{k,\textrm{fake-}e}^{\FakesCR})  \nonumber\\
\times~  & \prod_{j\in\mathrm{syst.}}\mathcal{G}(0|\theta_j)
\label{eq:likelihood_XS}
\end{align}
The EW \Zg{} signal contribution entering the \ZrcCR is taken into account in the statistical analysis. The fake-electron background in the CR bin at high \MET{} is determined by the product of $B_{\textrm{fake-}e}$ and the transfer factor $\mathcal{R}_{\textrm{fake-}e}$ (see Section~\ref{subsec:background}). The symbol $B_{\textrm{fake-}e}^{\FakesCR}$ represents the number of events with a fake electron in the corresponding bin of the \FakesCR{} at low \MET{}. For this measurement, in addition to the parameter of interest \muZgEWK, the normalization factors $\beta_{\Zboson\gamma_{\mathrm{strong}}}$ and $\beta_{\Wboson\gamma}$ corresponding to the strong-production \Zg{} component and, inclusively, to the \Wg{} component of the investigated distribution respectively, are allowed to float in the fit. 
 
The result of the maximum-likelihood fit to data in the 4 SR bins and 16 CR bins is shown in \cref{fig:srcr_postfit_mjj}, with the best-fit model propagated in all the regions. The SR bin-by-bin yields and CR yields are shown in \cref{tab:unyieldspostfitSepBinsmjj} for the SM process contributions and data. The level of agreement between the data and the prediction from background simulation is good overall and is better after the fit, as shown in the lower panel of \cref{fig:srcr_postfit_mjj}.
 
\begin{figure}[!htbp]
\centering
\includegraphics[width=0.8\linewidth]{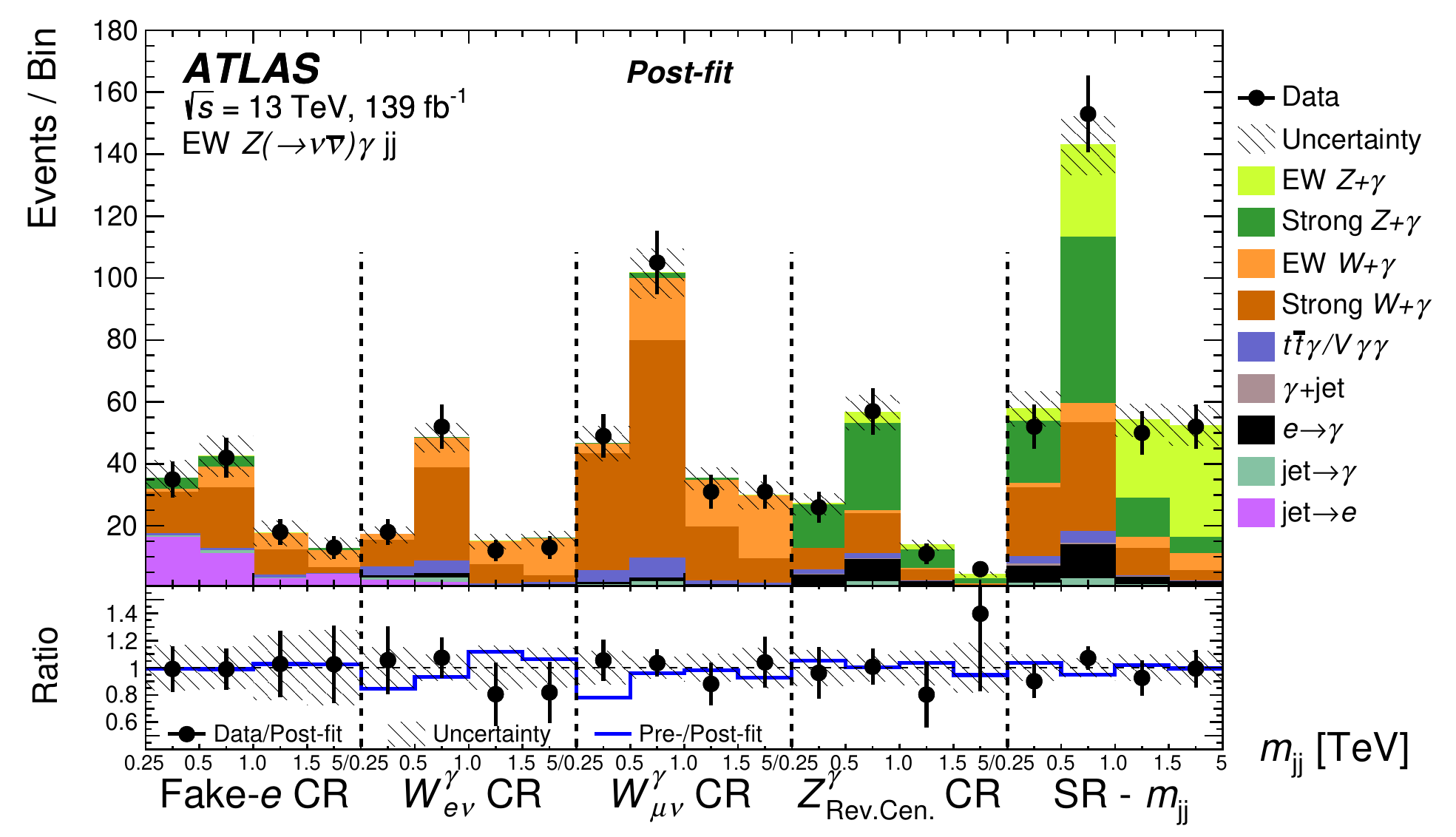}
\caption{Post-fit results for all \mjj SR and CR bins in the EW \Zg{} cross-section measurement as defined in Eq.~\eqref{eq:likelihood_XS} with the \muZgEWK signal normalization floating. The post-fit uncertainties include statistical, experimental, and theory contributions. The lower panel shows the ratio of data to the sum of all the SM process predictions and also compares the pre-fit and post-fit predictions.}
\label{fig:srcr_postfit_mjj}
\end{figure}
 
\begin{table}[!htbp]
\caption{Data yields and fitted predictions, after the fit to \integLumi of data with the \muZgEWK signal normalization floating as defined in \cref{eq:likelihood_XS}, for the four \mjj{} bins of the SR and the inclusive CRs. The uncertainties in the SM processes are derived by the fit and include the effects of nuisance parameter constraints and the correlation of systematic uncertainties. The individual uncertainties are correlated and do not necessarily add in quadrature to equal the total background uncertainty. A dash `--' indicates less than 0.01 events. 
}
\label{tab:unyieldspostfitSepBinsmjj}
\centering
\resizebox{0.99\textwidth}{!}{
\begin{tabular}{c|cccc|cccc}
\toprule
\multirow{ 2}{*}{Process} & \multirow{ 2}{*}{\FakesCR{}} & \multirow{ 2}{*}{\WeCR} & \multirow{ 2}{*}{\WmuCR} &\multirow{ 2}{*}{\ZrcCR} &  \multicolumn{ 4}{c}{SR - \mjj{} [\TeV{}]}   \\
& & & &  & 0.25-0.5 & 0.5-1.0 & 1.0-1.5 & $\geq1.5$ \\
\hline
Strong \Zg{} &  $8 \pm 8$ &  $0 \pm 1$ &  $3 \pm 2$ &  $50 \pm 12$ &  $20 \pm 6$~~ &  $54 \pm 12$ &  $13 \pm 5$~~ &  $5 \pm 2$ \\
EW \Zg{} &  $0.6 \pm 0.2$ &  $0.3 \pm  0.2$ &  $0.4 \pm 0.2$ &  $7 \pm 2$ &  $4 \pm 1$ &  $30 \pm 7$~~ &  $25 \pm 5$~~ &  $36 \pm 7$~~ \\
Strong \Wg{} &  $43 \pm 9$~~ &  $47 \pm 9$~~ &  $133 \pm 21$~~ &  $24 \pm 6$~~ &  $22 \pm 6$~~ &  $35 \pm 10$ &  $9 \pm 3$ &  $3 \pm 1$ \\
EW \Wg{} &  $19 \pm 6$~~ &  $31 \pm 7$~~ &  $59 \pm 13$ &  $1.4 \pm 0.5$ &  $2 \pm 1$ &  $6 \pm 1$ &  $4 \pm 1$ &  $5 \pm 1$ \\
jet$\rightarrow\gamma$ &  $1 \pm 1$ &  $2 \pm 2$ &  $3 \pm 2$ &  $2 \pm 2$ &  $1 \pm 1$ &  $2 \pm 2$ &  $1 \pm 1$ &  $0.4 \pm 0.3$ \\
jet$\rightarrow e$ &  $34 \pm 17$ &  $5 \pm 3$ &  -- &  -- &  -- &  -- &  -- &  -- \\
$e\rightarrow\gamma$ &  -- &  $2.7 \pm 0.4$ &  $2.9 \pm 0.4$ &  $13 \pm 1$~~ &  $6 \pm 1$ &  $11 \pm 1$~~ &  $2.6 \pm 0.4$ &  $1.4 \pm 0.3$ \\
\gjet{} &  -- &  -- &  -- &  $0.7 \pm 0.5$  &  $0.7 \pm 0.5$ &  $0.4 \pm 0.3$ &  $0.1 \pm 0.1$ &  $0.1 \pm 0.1$ \\
$\ttbar{}\gamma/V\gamma\gamma$ &  $3 \pm 1$ &  $9 \pm 2$ &  $13 \pm 2$~~ &  $3 \pm 1$ &  $2 \pm 1$ &  $4 \pm 1$ &  $0.4 \pm 0.2$ &  $0.1 \pm 0.1$ \\\hline
Fitted Yields &  $108 \pm 10$~~ &  $96 \pm 8$~~ &  $213 \pm 14$~~ &  $102 \pm 9$~~~~ &  $58 \pm 6$~~ &  $143 \pm 12$~~ &  $54 \pm 5$~~ &  $52 \pm 6$~~ \\\hline
Data &  108 &  95 &  216 &  100 &  52 &  153 &  50 &  52\\ 
Data/Fit & $1.00 \pm 0.14$ & $0.99 \pm 0.12$ & $1.01 \pm 0.09$ & $0.98 \pm 0.13$ & $0.90 \pm 0.15$ & $1.07 \pm 0.11$ & $0.93 \pm 0.16$ & $0.99 \pm 0.18$ \\\hline
\bottomrule
\end{tabular}}
\end{table}
 
The best-fit values of the three free-floating normalization factors appearing in Eq.~\eqref{eq:likelihood_XS} are reported in \cref{tab:postFitNF}.
In particular, the EW \Zg{} normalization best-fit value is \NormZgEWKDetail{}. The excess over the background-only hypothesis is quantified by a $p$-value using the profile likelihood ratio, evaluated at $\muZgEWK = 0$, as a test statistic. EW \Zgjj{} is observed with a significance of \ObsSigZg with respect to the other SM background processes, and the expected significance is \ExpSigZg. The statistical component of the uncertainty includes only the data statistical uncertainty with other sources such as the limited number of simulated events and the normalization of the backgrounds included in the systematic component of the uncertainty.
 
\begin{table}[!htbp]
\caption{The best-fit values and corresponding uncertainties of the three free-floating normalization factors derived from the statistical analysis described by Eq.~\eqref{eq:likelihood_XS}.}
\label{tab:postFitNF}
\centering
\begin{tabular}{c|c|c}
\toprule
$\muZgEWK$ & $\beta_{\Zboson\gamma_{\mathrm{strong}}}$ & $\beta_{\Wboson\gamma}$ \\
\hline
\NormZgEWK & \NormZgQCD & \NormWg \\
\bottomrule
\end{tabular}
\end{table}
 
The impact on the measurement of \muZgEWK from different groups of uncertainties is shown in Table~\ref{tab:LimitKerasImpact}. It is evaluated by repeating the fit procedure, after fixing the nuisance parameters corresponding to each group of systematic uncertainties, in turn, to their best-fit values, and subtracting the new variance ($\sigma^2$) of the best-fit value of \muZgEWK from the original variance. The data statistical uncertainty has the largest impact on the measured signal strength, followed by the signal acceptance uncertainties.
A small correlation is observed among the different sources of uncertainty. The signal uncertainties are divided into acceptance uncertainties for the signal events entering the fiducial volume and the shape uncertainties, which are the uncertainties in the shape of signal distributions within the fiducial volume. The signal acceptance uncertainties are assigned to the theoretical cross-section and not to the fiducial cross-section.
 
\begin{table}[!htbp]
\centering
\caption{The contributions from different groups of systematic uncertainties to the $\pm 1\sigma$ uncertainty bands of the \muZgEWK best-fit value and on \Binv and \Byyd 95\% CL limits.
The evaluation is performed by fixing a given group of systematic uncertainties to their best-fit values and subtracting the new variance ($\sigma^2$) of the best-fit value or the limit from the nominal variance including all systematic uncertainties. Due to residual correlations between categories, the sum in quadrature of the systematic uncertainties can differ from the actual value.
The uncertainty due to the finite number of data events (`Data stats.') is obtained by fixing all systematic uncertainties to their best-fit values.
The sum of all systematic uncertainties is estimated by subtracting the statistical variance component from the total variance. The experimental uncertainties and the uncertainty related to the size of MC simulated samples (`MC stats.') are treated as separate categories. The \Vg{} theory entry includes the theoretical uncertainties in strong \Zg{}, EW \Wg{} and strong \Wg{} production for \muZgEWK; however, for \Binv and \Byyd, it also includes those from EW \Zg{}.
For the last two columns the impact of systematic uncertainties is computed from a fit to data with $\Binv = 0$ or $\Byyd = 0$ for each respective column.}
\label{tab:LimitKerasImpact}
\resizebox{0.9\textwidth}{!}{
\begin{tabular}{c|ccc}
\toprule
Source & 1$\sigma$ Uncertainty on \muZgEWK & 1$\sigma$ Uncertainty on \Binv &   1$\sigma$ Uncertainty on \Byyd \\
\hline
Jet scale and resolution	 &	0.076 &	0.045 &	0.0011\\
\Vg{} theory			 &	0.067 &	0.044 &	0.0018\\
\pileup 				&	0.040 &	0.021 &	0.0004 \\
Photon	      			 &	0.035 &	0.031 &	0.0011\\
$e\rightarrow\gamma$, jet$\rightarrow e,\gamma$ Bkg. &	0.035 & 0.034 &	0.0028 \\
Lepton				&	0.027 &	0.003 &	0.0008 \\
\met				&	0.023 &	0.018 &	0.0003 \\\hline
Signal theory shape		&	0.020 &	-- &	-- \\
Signal theory acceptance	&	0.12~~ &	-- &	-- \\  \hline
Data stats.			 &	0.16~~ &	0.11~~ &	0.0056\\
\Wg/\Zg Norm.  	      		&	0.073 &	0.013 &	0.0004 \\
MC stats.			 &	0.063 &	0.046 &	0.0026\\\hline
Total				&	0.25~~ & 0.15~~ &	0.0073 \\
\bottomrule
\end{tabular} }
\end{table}
 
The measured fiducial cross-section is extracted by taking the product of the signal strength, \muZgEWK, and the predicted cross-section times branching ratio to neutrinos in the fiducial volume defined in Section~\ref{subsec:mjjbinning}. The measurement and SM prediction agree within the measurement uncertainties. The measured fiducial cross-section is $\sigma^{\text{fid.}}_{Z(\rightarrow \nu\nu)\gamma\text{EW}} = \FidZgEWKDetail$, which includes the contribution from the interference term with the strong production of \Zg{}. The interference computed through \MADGRAPH{} is 2\% in the fiducial volume and is treated as an uncertainty in the EW \Zg{} cross-section. The theoretical \MADGRAPH{} cross-section including the 0.3\% NLO QCD $K$-factor correction from VBFNLO is \MGXSZgEWKDetail{} = \MGXSZgEWK{}. The jet-veto is not part of the fiducial phase-space definition; the loss in efficiency in simulation for this veto is 5\%.
 
\subsection{Fit model and results for \Hinv{} search}
\label{subsec:invfitmodel}
 
In the search for \Hinv{}, events are categorized into four bins according to the DNN output score. These bins enter the likelihood function definition in \cref{eq:likelihood}. In addition to the SR bins, the correspondingly binned \WeCR, \WmuCR, \ZrcCR, and \FakesCR are included in the likelihood function definition to provide constraints on the background contribution to the SR. The signal contribution in the \ZrcCR is taken into account in the statistical analysis.
In the likelihood function definition, all the normalization factors $\beta$ from Eq.~\eqref{eq:likelihood_XS} are fixed to one except the ones corresponding to the \Wg{} and fake-$e$ background contributions. The $\muZgEWK$ normalization factor is fixed to one because the shape of this contribution and the one of the \Hinv{} signal are so similar that leaving the former unconstrained would largely affect the search sensitivity. The result of the EW \Zg{} cross-section measurement described in Section~\ref{subsec:zgewkfitmodel} further supports this assumption.
The $\beta_{\Zboson\gamma_{\mathrm{strong}}}$ normalization factor is fixed to one because the CR yields are not large enough to reduce the theoretical uncertainties. In addition, the observed normalization in the EW \Zg{} cross-section measurement is consistent with unity in Table~\ref{tab:postFitNF}. The likelihood is
\begin{align}
& \mathcal{L}(\mu,\beta_{\Wboson\gamma},\beta_{\textrm{fake-}e},\overrightarrow{\theta}) & = & \prod_{i\in \mathrm{SR~bins}}\mathcal{P}(N_i|\beta_{\Wboson\gamma} B_{i,\Wboson\gamma} + B_{i,\Zboson\gamma} + B_{i,\gj} + B_{i,\mathrm{others}} + \mu S_i)  \nonumber \\
&             & \times~  & \prod_{k\in \mathrm{CR~bins}}\mathcal{P}(N_k^{\WeCR}|\beta_{\Wboson\gamma} B_{k,\Wboson\gamma}^{\WeCR} + B_{k,\textrm{non-\Wboson}}^{\WeCR} + \beta_{k,\mathrm{fake-}e} B_{k,\textrm{fake-e}}^{\FakesCR} \mathcal{R}_{\textrm{fake-}e}) \nonumber\\
&             & \times~  & \prod_{k\in \mathrm{CR~bins}}\mathcal{P}(N_k^{\WmuCR}|\beta_{\Wboson\gamma} B_{k,\Wboson\gamma}^{\WmuCR} + B_{k,\textrm{non-\Wboson}}^{\WmuCR}) \nonumber\\
&             & \times~  & \prod_{k\in \mathrm{CR~bins-1}}\mathcal{P}(N_k^{\ZrcCR}|B_{k,\mathrm{TOT.}}^{\ZrcCR} + \mu S_k^{\ZrcCR}) \nonumber\\
&             & \times~  & \prod_{k\in \mathrm{CR~bins}}\mathcal{P}(N_k^{\FakesCR}|\beta_{\Wboson\gamma} B_{k,\Wboson\gamma}^{\FakesCR} + B_{k,\textrm{non-\Wboson}}^{\FakesCR} + \beta_{k,\textrm{fake-}e} B_{k,\textrm{fake-e}}^{\FakesCR}) \nonumber\\
&             & \times~  & \prod_{j\in\mathrm{syst.}}\mathcal{G}(0|\theta_j)~\mathrm{,}
\label{eq:likelihood_Hinv}
\end{align}
where $B_{k,\textrm{TOT.}}^{\ZrcCR}$ is the sum of the expected background yields in the bin $k$ of the \ZrcCR{}, and $B_{k,\Wboson\gamma}$ is the sum of the EW and strong \Wg{} events with their ratios fixed from theoretical predictions. Because of a very small yield in the fourth bin of the \ZrcCR{} for the \Binv{} search, the third and fourth bins are merged into one bin covering 0.6--1.0 in DNN output score.
The result of the maximum-likelihood fit to data in the 4 SR and 15 CR bins is shown in \cref{fig:srcr_postfit_mva}, with the best-fit model propagated in all the regions. The SR bin-by-bin yields and CR yields are shown in \cref{tab:unyieldspostfitSepBinsKeras} for the background contribution, a benchmark \Hinv signal contribution, and recorded data. The fitted normalization of the sum of the EW and strong \Wg{} events relative to the SM prediction is $1.07 \pm 0.18$; the fake-electron normalization is not reported because no comparison with the SM predictions is possible. The level of agreement between the data and the prediction from background simulation is good overall, and is better after the fit, as shown in the lower panel of \cref{fig:srcr_postfit_mva}. No evidence of a new physics contribution is visible on top of the background prediction. The observed (expected) upper limit on \Binv is \ObsHinvBrLimit (\ExpHinvBrLimit) at 95\% CL. The impact on the limit from different groups of uncertainties is shown in Table~\ref{tab:LimitKerasImpact}, and the results are evaluated in the same way as for observation of EW \Zg{}, described in \cref{subsec:zgewkfitmodel}, except that \Binv{} is fixed to zero.
 
\begin{figure}[!htbp]
\centering
\includegraphics[width=0.8\linewidth]{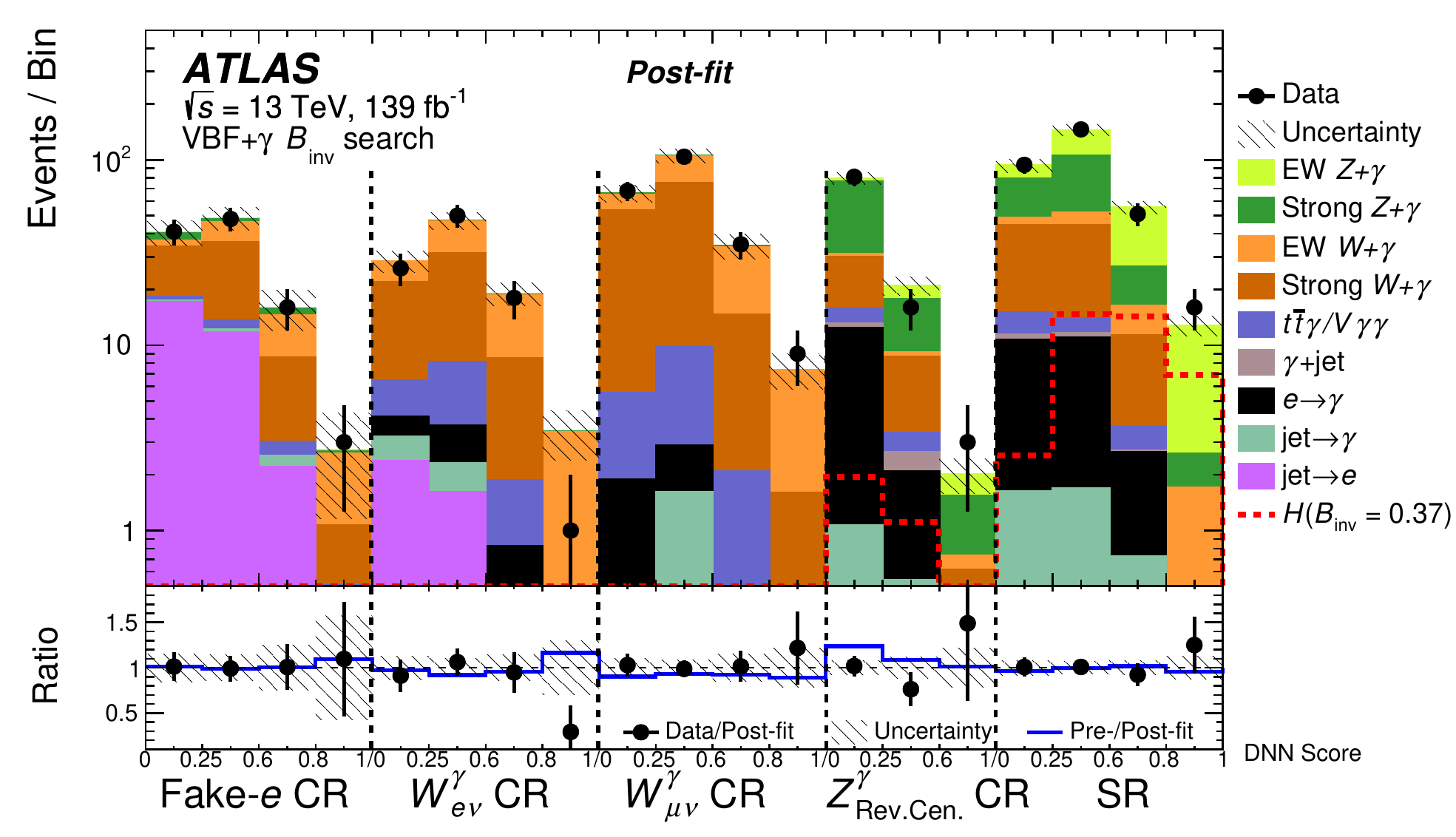}
\caption{Post-fit results for all DNN SR and CR bins in the search for \Hinv{} as defined in Eq.~\eqref{eq:likelihood_Hinv} with the \Binv{} signal normalization set to zero. For the \ZrcCR{}, the third bin contains all events with DNN output score values of 0.6--1.0.
The \Hinv{} signal is scaled to a \Binv{} of 37\%. The post-fit uncertainties include statistical, experimental, and theoretical contributions. The lower panel shows the ratio of data to the post-fit yields and also compares the pre-fit and post-fit background predictions.}
\label{fig:srcr_postfit_mva}
\end{figure}
 
\begin{table}[!htbp]
\caption{Data yields and background predictions, after the fit to \integLumi of data with the \Binv{} signal normalization set to zero, for the four DNN output score bins of the SR and the inclusive CRs. The uncertainties in the backgrounds are derived by the fit and include the effects of nuisance parameter constraints and the correlation of systematic uncertainties; as a result, the uncertainties in the total background can be smaller than the sum of those in the single contributions. The predicted signal yields, assuming a 37\% branching ratio for \Hinv{} are shown with their associated uncertainties. A dash `--' indicates less than 0.01 events.
}
\label{tab:unyieldspostfitSepBinsKeras}
\centering
\resizebox{0.99\textwidth}{!}{
\begin{tabular}{c|cccc|cccc}
\toprule
\multirow{ 2}{*}{Process}  & \multirow{ 2}{*}{\FakesCR{}} & \multirow{ 2}{*}{\WeCR} & \multirow{ 2}{*}{\WmuCR} &\multirow{ 2}{*}{\ZrcCR} & \multicolumn{ 4}{c}{SR - DNN output score}  \\
& & &  & & 0-0.25 & 0.25-0.6 & 0.6-0.8 & 0.8-1.0 \\
\hline
Strong \Zg{} &   $7 \pm 4$ & $0.3 \pm 0.1$ & $3 \pm 1$ & $55 \pm 6$~~ & $31 \pm 6$~~ &  $53 \pm 10$ &  $10 \pm 2$~~ &  $0.9 \pm 0.2$ \\
EW \Zg{}&  $0.6 \pm 0.1$ & $0.3 \pm 0.1$ & $0.4 \pm 0.2$ & $7 \pm 1$ &  $14 \pm 2$~~ &  $39 \pm 5$~~ &  $29 \pm 4$~~ &  $10 \pm 2$~~ \\
Strong \Wg{} &  $44 \pm 7$~~ & ~$46 \pm 6.1$ & $127 \pm 13$~~ & $20 \pm 2$~~ & $29 \pm 6$~~ &  $31 \pm 9$~~ &  $8 \pm 2$ &  $0.1 \pm 0.2$ \\
EW \Wg{}  &  $21 \pm 3$~~ & $35 \pm 4$~~ & $66 \pm 7$~~ & $1.5 \pm 0.2$ & $5 \pm 1$ &  $8 \pm 1$ &  $5 \pm 1$ &  $1.2 \pm 0.3$ \\
jet$\rightarrow\gamma$ &  $1 \pm 1$ & $2 \pm 1$ & $2 \pm 2$ & $2 \pm 1$ &  $2 \pm 2$ &  $2 \pm 2$ &  $0.7 \pm 0.8$ & -- \\
jet$\rightarrow e$ &  $32 \pm 15$  &  $4 \pm 2$ &  -- &  -- &  -- &  -- &  -- & -- \\
$e\rightarrow\gamma$ &  -- &  $2.6 \pm 0.2$ &  $2.9 \pm 0.2$ &  $13 \pm 1$~~ & $9 \pm 1$ &  $9 \pm 1$ &  $1.9 \pm 0.2$ &  $0.3 \pm 0.1$ \\
\gjet{} & -- & -- & -- & $1.4 \pm 0.5$  & $0.7 \pm 0.5$ &  $0.6 \pm 0.4$ &  $0.04 \pm 0.03$ & -- \\
$\ttbar{}\gamma/V\gamma\gamma$ & $3 \pm 1$ & $8 \pm 1$ & $12 \pm 2$~~ & $3 \pm 1$ &  $4 \pm 1$ &  $2.4 \pm 0.4$ &  $0.9 \pm 0.2$ &  $0.01 \pm 0.01$ \\\hline
Fitted Bkg  &  $108 \pm 10$~~ &  $97 \pm 7$~~  & $214 \pm 14$~~ & $101 \pm 7$~~~~ &   $93 \pm 8$~~ &  $145 \pm 11$~~ &  $55 \pm 5$~~ &  $13 \pm 2$~~ \\\hline
$H$ ($\Binv = 0.37$)  & -- & -- & -- & $3.3 \pm 0.4$ &  $2.6 \pm 0.3$ &  $15 \pm 2$ &  $14 \pm 2$ &  $7 \pm 1$ \\ \hline
Data &  108 & 95 & 216 & 100 &  94 &  146 &  51 &  16 \\
Data/Fit & $1.01 \pm 0.14$ & $0.97 \pm 0.12$ & $1.01 \pm 0.09$ & $0.98 \pm 0.12$ & $1.01 \pm 0.13$ & $1.01 \pm 0.11$ & $0.92 \pm 0.15$ & $1.25 \pm 0.35$ \\
\bottomrule
\end{tabular}   }
\end{table}
 
\subsection{Fit model and results for \Hyyd{} search}
\label{subsec:ggdfitmodel}
 
In the search for a Higgs boson decaying into a $\gamma\gammad$ pair, the most powerful discriminating observable is the photon--\met transverse mass \mtgammet, so this observable is used to search for this new physics signal. The events entering the dedicated SR$_{\gammad}$ (see \cref{sec:event_selection}) are separated into five \MT{} bins, as described in \cref{subsec:mtbinning}. Because the relative contributions of \Hyyd signal produced through ggF and VBF production vary with \mjj{}, events are also separated into two \mjj{} categories, those with $\mjj< 1$~\TeV{} and those with $\mjj\geq 1$~\TeV{}.
A total of ten bins in the SR as well as ten bins in each CR enter the likelihood function definition, which is equivalent to the one in \cref{eq:likelihood_Hinv} other than having a different number of bins in the SR and CRs and a different signal benchmark model for the interpretation in the context of the \Hyyd search. The fixing of the $\beta$ and $\muZgEWK$ normalization factors to unity in Eq.~\eqref{eq:likelihood_Hinv} is repeated to be consistent with Section~\ref{subsec:invfitmodel}, but allowing them to float in the fit does not change the results within rounding.
The SR bin-by-bin yields and CR yields are shown in \cref{tab:unyieldspostfitSepBinsMT10} for the background contribution and recorded data yields after a fit to the background-only contributions.
 
\begin{sidewaystable}
\caption{Event yields of data and background predictions, after the fit to \integLumi of data with the \Byyd{} signal normalization set to 0, for the ten bins of the SR$_{\gammad}$ selection based on [\MT,\mjj{}] binning, and for the inclusive CRs. The uncertainties in the backgrounds are derived by the fit and include the effects of nuisance parameter constraints and the correlation of systematic uncertainties; as a result, the uncertainties in the total background can be smaller than the sum of those in the single contributions. The predicted signal yields, assuming the SM production cross-section for a 125 \GeV{} Higgs boson and $\Byyd = 0.02$, are shown with their associated uncertainties. A dash `--' indicates less than 0.01 events.
}
\label{tab:unyieldspostfitSepBinsMT10}
\centering
\resizebox{\textwidth}{!}{
\begin{tabular}{c|cccc|ccccc|ccccc}
\toprule
\multirow{ 3}{*}{Process} & \multirow{ 3}{*}{\FakesCR{}} & \multirow{ 3}{*}{\WeCR} & \multirow{ 3}{*}{\WmuCR} &\multirow{ 3}{*}{\ZrcCR} &  \multicolumn{ 5}{c}{SR - $\mjj{}<1~\TeV{}$} &  \multicolumn{ 5}{c}{SR - $\mjj{}\geq1~\TeV{}$} \\
& & & & &  \multicolumn{ 5}{c}{\MT{} [\GeV{}]} &  \multicolumn{ 5}{c}{\MT{} [\GeV{}]}   \\
& & & & & 0-90 & 90-130 & 130-200 & 200-350 & $\geq350$& 0-90 & 90-130 & 130-200 & 200-350 & $\geq350$\\
\hline
Strong \Zg{} &  $38 \pm 6$~~ &  $3 \pm 2$ &  $8 \pm 2$ &  $73 \pm 12$ &  $29 \pm 7$~~ &  $30 \pm 7$~~ &  $29 \pm 7$~~ &  $22 \pm 5$~~ &  $5 \pm 1$ &  $6 \pm 2$ &  $6 \pm 2$ &  $7 \pm 2$ &  $4 \pm 1$ &  $2 \pm 1$ \\
EW \Zg &  $7 \pm 1$ &  $0.9 \pm 0.2$ &  $0.8 \pm 0.2$ &  $13 \pm 2$~~ &  $10 \pm 2$~~ &  $9 \pm 1$ &  $14 \pm 2$~~ &  $12 \pm 2$~~ &  $4 \pm 1$ &  $23 \pm 3$~~ &  $14 \pm 2$~~ &  $25 \pm 3$~~ &  $20 \pm 3$~~ &  $9 \pm 2$ \\
Strong \Wg &  $105 \pm 18$~~ &  $126 \pm 17$~~ &  $244 \pm 30$~~ &  $115 \pm 16$~~ &  $117 \pm 15$~~ &  $22 \pm 4$~~ &  $27 \pm 5$~~ &  $17 \pm 3$~~ &  $3 \pm 1$ &  $34 \pm 5$~~ &  $8 \pm 1$ &  $5 \pm 2$ &  $2 \pm 1$ &  $2 \pm 1$ \\
EW \Wg &  $37 \pm 9$~~ &  $62 \pm 14$ &  $114 \pm 24$~~ &  $5 \pm 1$ &  $16 \pm 3$~~ &  $4 \pm 1$ &  $4 \pm 1$ &  $2.1 \pm 0.4$ &  $0.5 \pm 0.2$ &  $31 \pm 5$~~ &  $3 \pm 1$ &  $4 \pm 1$ &  $3 \pm 1$ &  $1.5 \pm 0.3$ \\
jet$\rightarrow\gamma$ &  $4 \pm 3$ &  $5 \pm 4$ &  $6 \pm 5$ &  $5 \pm 4$ &  $10 \pm 8$~~ &  $3 \pm 2$ &  $1 \pm 1$ & $1 \pm 1$ &  $0.1 \pm 0.1$ &  $3 \pm 3$ &  $0.5 \pm 0.5$ &  $0.6 \pm 0.5$ &  $0.1 \pm 0.1$  &  $0.1 \pm 0.1$ \\
jet$\rightarrow e$  &  $46 \pm 21$ &  $6 \pm 2$ &  -- &  -- &  -- &  -- &  -- &  -- &  -- &  -- &  -- &  -- &  -- &  -- \\
$e\rightarrow\gamma$ &  -- &  $5 \pm 1$ &  $6 \pm 1$ &  $160 \pm 14$~~ &  $179 \pm 15$~~ &  $20 \pm 2$~~ &  $5 \pm 1$ &  $2.8 \pm 0.3$ & $0.5 \pm 0.1$ &  $76 \pm 6$~~ &  $6 \pm 1$ &  $1.4 \pm 0.1$ &  $0.8 \pm 0.1$ &  $0.2 \pm 0.1$\\
\gjet{} &  -- &  -- &  -- &  $21 \pm 6$~~ &  $7 \pm 7$ &  $6 \pm 6$ &  $2 \pm 2$ &  $0.2 \pm 0.2$ &  $0.1 \pm 0.1$ &  $2 \pm 1$ &  $1 \pm 1$ &  $0.4 \pm 0.4$ &  $0.1 \pm 0.1$ &  $0.03 \pm 0.02$\\
$\ttbar{}\gamma/V\gamma\gamma$ &  $7 \pm 2$ &  $25 \pm 4$~~ &  $32 \pm 4$~~ &  $9 \pm 1$ &  $15 \pm 2$~~ &  $4 \pm 1$ &  $2.7 \pm 0.3$ &  $2.3 \pm 0.3$ &  $0.1 \pm 0.1$ &  $9 \pm 1$ &  $0.8 \pm 0.3$ &  $0.1 \pm 0.1$ &  $0.04 \pm 0.09$ &  $0.03 \pm 0.01$ \\\hline
Fitted Bkg &  $245 \pm 16$~~ &  $233 \pm 11$~~ &  $411 \pm 18$~~ &  $398 \pm 19$~~ &   $363 \pm 15$~~ &  $97 \pm 8$~~ &  $84 \pm 7$~~ &  $59 \pm 5$~~ &  $12 \pm 2$~~ &  $184 \pm 8$~~~~ &  $40 \pm 3$~~ &  $44 \pm 3$~~ &  $30 \pm 3$~~ &  $14 \pm 2$~~ \\\hline
VBF $H$125 ($\Byyd = 0.02$)  & -- & -- & -- & $0.27 \pm 0.03$ &  $3.6 \pm 0.4$ &  $8.8 \pm 1.1$ &  $1.7 \pm 0.2$ &  -- & -- & $5.7 \pm 0.7$  & $12.2 \pm 1.4$~~  & $2.7 \pm 0.3$ &  -- &  --\\
ggF $H$125 ($\Byyd = 0.02$)  & -- & -- & -- & $0.4 \pm 0.1$ &  $1.9 \pm 0.6$ &  $4.3 \pm 1.3$ &  $0.8 \pm 0.3$ &  -- & -- & $0.7 \pm 0.3$  & $1.2 \pm 0.4$  & $0.2 \pm 0.1$ &  -- &  --\\\hline
Data &  244 &  236 &  410 &  348 &  362 &  112 &  75 &  59&  8 &  188 &  31 &  56 &  35 &  9 \\
Data/Fit & $1.00 \pm 0.09$ & $1.01 \pm 0.08$ & $1.00 \pm 0.07$ & $0.88 \pm 0.12$ & $1.00 \pm 0.07$ & $1.15 \pm 0.14$ & $0.89 \pm 0.13$ & $1.00 \pm 0.16$ & $0.65 \pm 0.25$ & $1.02 \pm 0.09$ & $0.78 \pm 0.15$ & $1.27 \pm 0.19$ & $1.17 \pm 0.23$ & $0.63 \pm 0.23$ \\
\bottomrule
\end{tabular}}
\end{sidewaystable}
 
The result of the maximum-likelihood fit with the \Byyd{} signal normalization set to zero in the ten SR bins and four inclusive CRs is shown in \cref{fig:srcr_postfit_SRonly_mt10}. The CRs are shown inclusively to reduce the number of bins presented, but the same binning as the SR is used in the fit model.
The data and predictions from background simulation agree within the reported uncertainties, apart from a small deficit in the data in the bins corresponding to the highest \MT{} values. The pre-fit background predictions in the highest \MT{} bins are pulled down by the fit to data, and uncertainties describing these differences, which increase in this high \MT{} range, come from the interference between EW and strong \Zgamma{} production as well as a comparison between \MGNLO{} and \SHERPA{} simulation for the strong-production \Zgamma{} background contributions. Overall, the level of agreement is better after the fit, as shown in the lower panel of \cref{fig:srcr_postfit_SRonly_mt10}; no evidence of a new physics contribution is visible on top of the background prediction.
 
\begin{figure}
\centering
\includegraphics[width=0.8\linewidth]{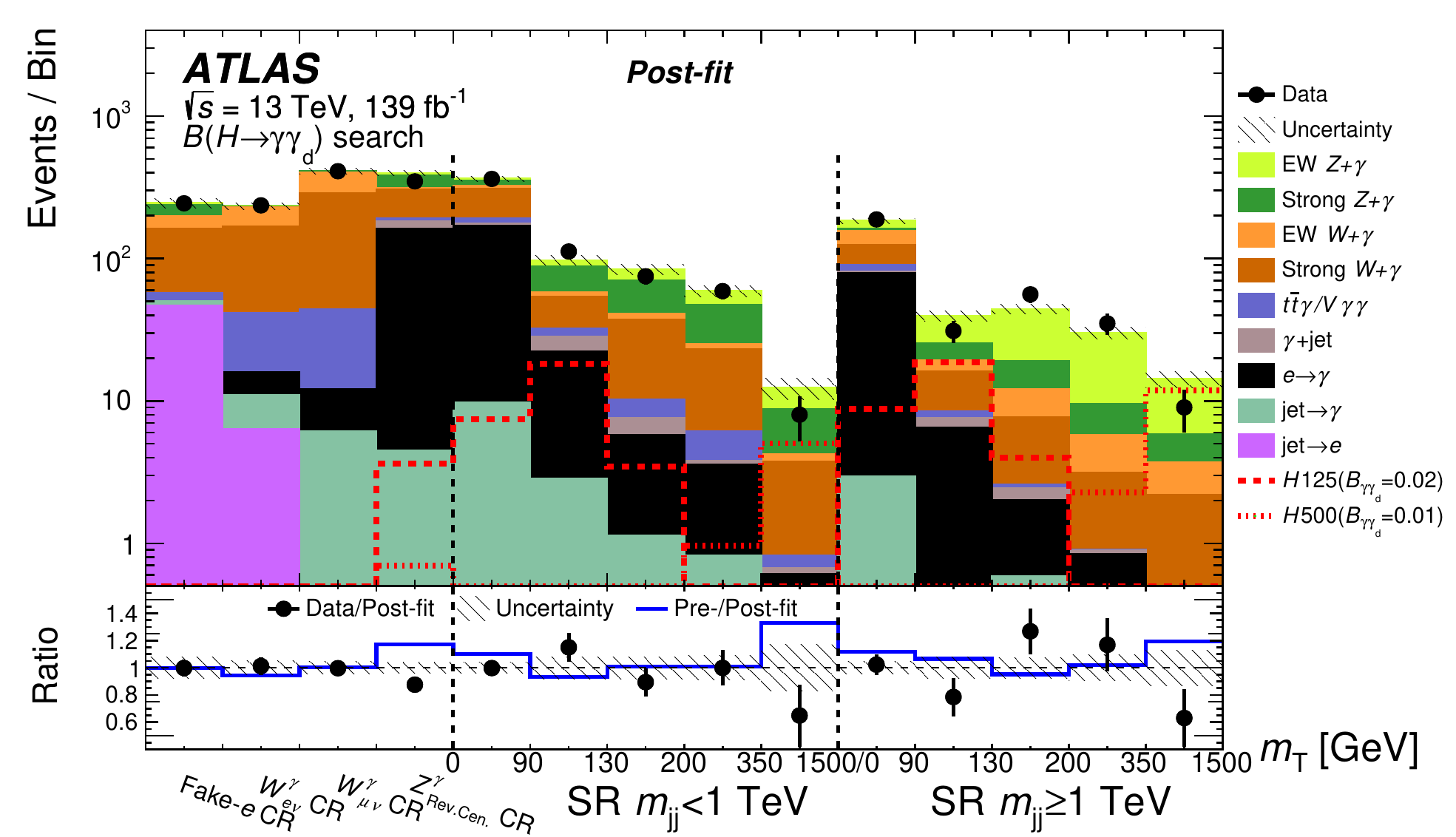}
\caption{Post-fit results for the ten [\mjj{},\MT{}] bins constituting the SR defined for the dark-photon search, and the CRs as defined in Eq.~\eqref{eq:likelihood_Hinv} with the \Byyd{} signal normalization set to zero. A \Hyyd signal is shown for two different mass hypotheses (125 \GeV, 500 \GeV) and scaled to a branching ratio of 2\% and 1\%, respectively. The post-fit uncertainties include statistical, experimental, and theoretical contributions. The lower panel shows the ratio of data to the sum of all the background contributions as well as a comparison of the pre-fit and post-fit background predictions.}
\label{fig:srcr_postfit_SRonly_mt10}
\end{figure}
 
\cref{fig:DP_kinematics} shows the distribution of \mtgammet in the inclusive SR$_{\gammad}$ (no \mjj split), and also shows the shape of a \Hyyd signal for two different mass hypotheses compared with same post-fit background predictions and data as for \cref{fig:srcr_postfit_SRonly_mt10}. Thus the reasons for the change in the pre-fit predictions for the highest \MT{} values are the same as described for \cref{fig:srcr_postfit_SRonly_mt10}. Good agreement between data and the background expectations in the \mtgammet distribution is observed also in the CRs considered in the statistical analysis.
 
\begin{figure}[!htbp]
\centering
\includegraphics[width=0.8\textwidth]{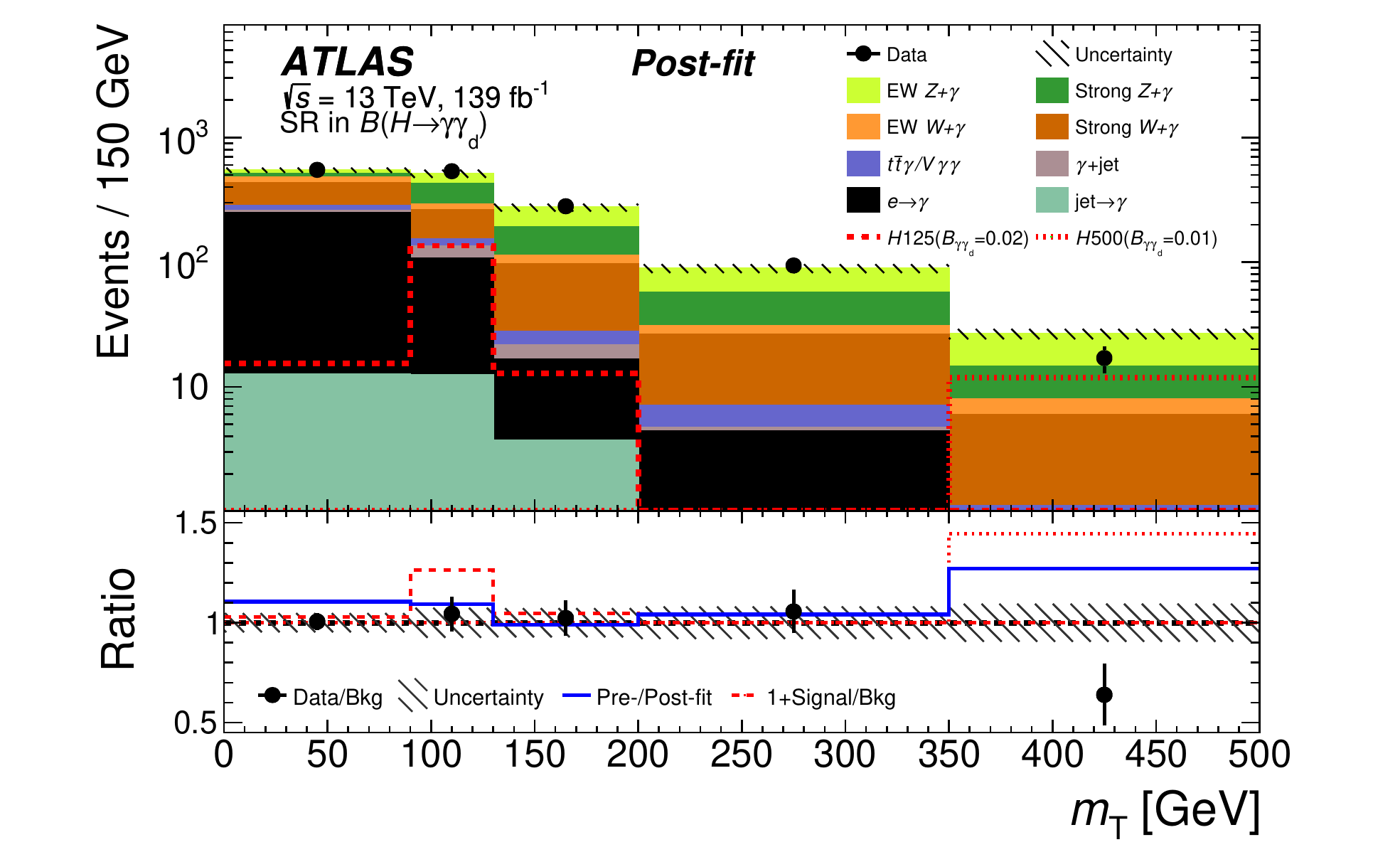}
\caption{Post-fit \mtgammet distribution in the inclusive signal region for the dark-photon search with the 125~\GeV{} mass \Byyd{} signal normalization set to zero. A \Hyyd signal is shown for two different mass hypotheses, 125 \GeV{} and 500 \GeV{}, and scaled to a \Byyd of 2\% and 1\%, respectively. The lower panel shows the ratio of data to the sum of all the background contributions, a comparison with the pre-fit background prediction, and the signal-to-background ratio shifted by 1.0 (to share the same vertical axis). Events with \mtgammet{} larger than the rightmost bin boundary are added to that bin.}
\label{fig:DP_kinematics}
\end{figure}
 
The statistical analysis sets an observed (expected) upper limit on \Byyd of \ObsDPBrLimit (\ExpDPBrLimit) at 95\% CL when considering both the VBF and ggF Higgs boson production mechanisms at a Higgs boson mass of 125~\GeV{}. If considering a BSM scalar boson with a mass of 125~\GeV{} produced through VBF, the observed (expected) upper limit on the cross-section times branching ratio is \ObsDPXSLimitVBF (\ExpDPXSLimitVBF) at 95\%. For such a BSM scalar boson produced through ggF, the observed (expected) upper limit on the cross-section times branching ratio is \ObsDPXSLimitggF (\ExpDPXSLimitggF) at 95\%, which shows that the sensitivity is dominated by the VBF production mode.
 
The 95\% CL limit on $\sigma^\text{VBF}\times\Byyd$ has also been calculated for a VBF-produced Higgs boson with different mass hypotheses in the narrow width approximation (NWA), ranging from 60~\GeV{} to 2~\TeV, as shown in \cref{fig:massVBFDarkPhLimitXS}.
The cross-section for a VBF-produced Higgs boson decreases rapidly with increasing boson mass, leading to smaller signal yields in the SR.
The signal corresponding to a high-mass scalar mediator peaks towards high values of \mtgammet, where the smaller background leads to good sensitivity despite the small expected signal.
 
\begin{figure*}[tbph!]
\begin{center}
\includegraphics[width=0.8\columnwidth]{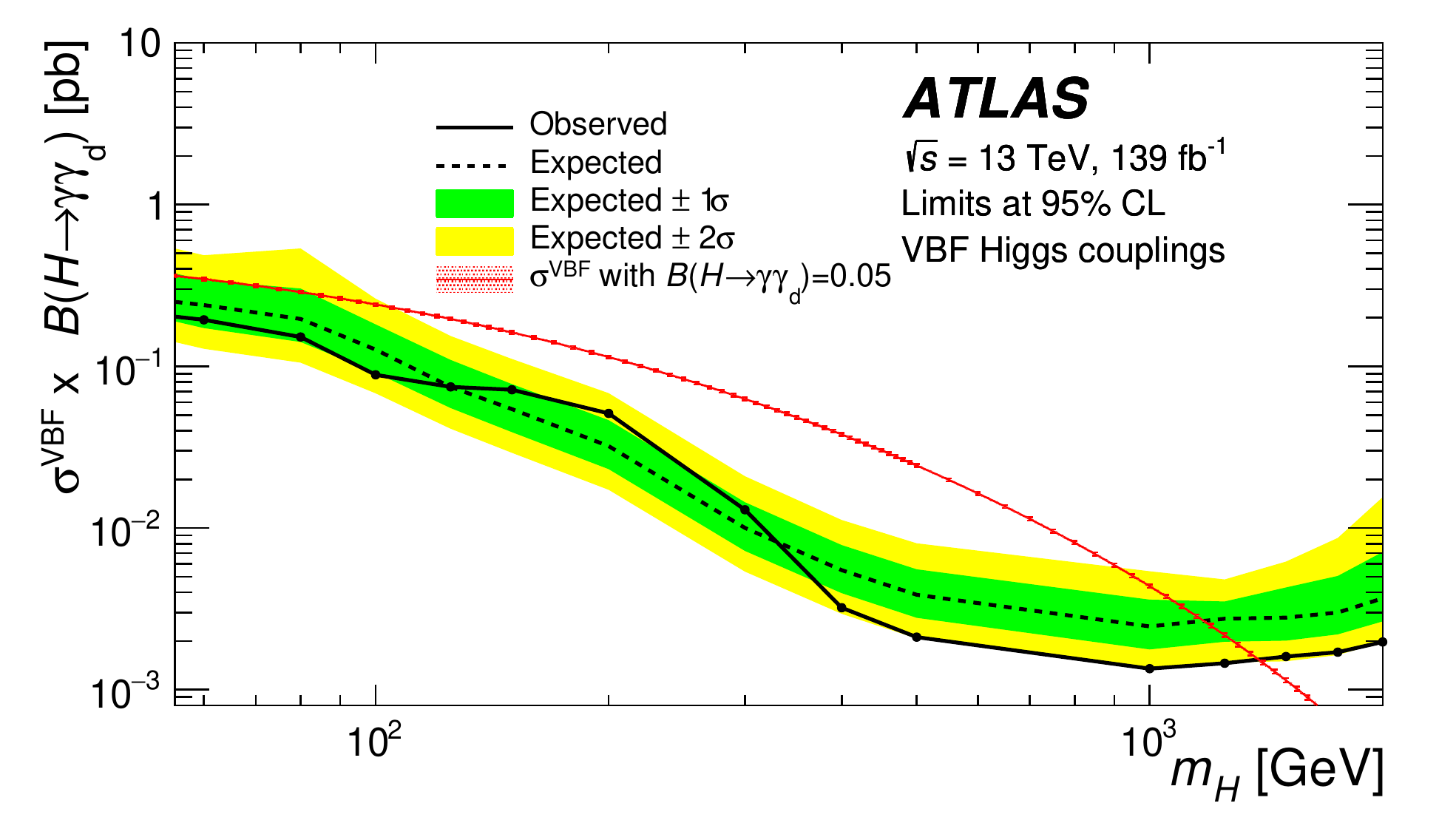}
\end{center}
\caption{The 95\% CL upper limit on the Higgs boson production cross-section times branching ratio to $\gamma\gammad$ for different VBF-produced scalar-mediator-mass hypotheses in the NWA. The theoretically predicted cross-section of a Higgs boson produced via VBF and with the \Byyd = 5\% is superimposed on the $\pm1\sigma$ and $\pm2\sigma$ NNLO QCD$+$NLO EW uncertainty bands of the expected production cross-section limit.}
\label{fig:massVBFDarkPhLimitXS}
\end{figure*}
 
The impact of various sources of uncertainty on the \Byyd upper limit is shown in Table~\ref{tab:LimitKerasImpact}, evaluated in the same way as for the \Hinv search, described in \cref{subsec:invfitmodel}. The statistical uncertainty of the yields of data events in SR$_{\gammad}$ has the largest impact on the limit determination. A negligible correlation is observed among the nuisance parameters corresponding to the different sources of uncertainty.

\section{Conclusion}
\label{sec:conclusion}

Data collected from \integLumi of 13 \TeV{} proton--proton collisions by the ATLAS experiment during the Run~2 of the LHC are scrutinized in a VBF-favoured signature of two forward jets, \MET{}, and a photon, to provide constraints on several SM and BSM processes.
The observation of SM EW \Zg{} production is reported with an observed (expected) significance of \ObsSigZg{} (\ExpSigZg). The fitted normalization for the EW \Zg{} process relative to the SM prediction is $\muZgEWK=\NormZgEWK{}$, corresponding to a measured cross-section of \FidZgEWK{} in the considered fiducial volume.
A search for Higgs bosons decaying solely into invisible particles is performed in the same final-state signature. Because no significant excess is observed, 95\% CL upper limits of \ObsHinvBrLimit (\ExpHinvBrLimit) are set on the observed (expected) branching ratio to invisible particles.
A search for Higgs bosons decaying into a photon and a dark photon is also performed, and the results exclude at 95\% CL cross-section times branching ratio values ranging from 0.15~pb for a scalar mediator with a mass of 60~\GeV{} to 3~fb for a scalar mediator with a mass of 2~\TeV{}.
For a Higgs boson mass of 125~\GeV{}, the observed (expected) 95\% CL upper limit on the Higgs boson branching ratio to $\gamma\gammad$ is \ObsDPBrLimit{} (\ExpDPBrLimit), the most stringent to date.

\section*{Acknowledgements}


We thank CERN for the very successful operation of the LHC, as well as the
support staff from our institutions without whom ATLAS could not be
operated efficiently.
 
We acknowledge the support of
ANPCyT, Argentina;
YerPhI, Armenia;
ARC, Australia;
BMWFW and FWF, Austria;
ANAS, Azerbaijan;
SSTC, Belarus;
CNPq and FAPESP, Brazil;
NSERC, NRC and CFI, Canada;
CERN;
ANID, Chile;
CAS, MOST and NSFC, China;
Minciencias, Colombia;
MSMT CR, MPO CR and VSC CR, Czech Republic;
DNRF and DNSRC, Denmark;
IN2P3-CNRS and CEA-DRF/IRFU, France;
SRNSFG, Georgia;
BMBF, HGF and MPG, Germany;
GSRI, Greece;
RGC and Hong Kong SAR, China;
ISF and Benoziyo Center, Israel;
INFN, Italy;
MEXT and JSPS, Japan;
CNRST, Morocco;
NWO, Netherlands;
RCN, Norway;
MEiN, Poland;
FCT, Portugal;
MNE/IFA, Romania;
JINR;
MES of Russia and NRC KI, Russian Federation;
MESTD, Serbia;
MSSR, Slovakia;
ARRS and MIZ\v{S}, Slovenia;
DSI/NRF, South Africa;
MICINN, Spain;
SRC and Wallenberg Foundation, Sweden;
SERI, SNSF and Cantons of Bern and Geneva, Switzerland;
MOST, Taiwan;
TAEK, Turkey;
STFC, United Kingdom;
DOE and NSF, United States of America.
In addition, individual groups and members have received support from
BCKDF, CANARIE, Compute Canada and CRC, Canada;
COST, ERC, ERDF, Horizon 2020 and Marie Sk{\l}odowska-Curie Actions, European Union;
Investissements d'Avenir Labex, Investissements d'Avenir Idex and ANR, France;
DFG and AvH Foundation, Germany;
Herakleitos, Thales and Aristeia programmes co-financed by EU-ESF and the Greek NSRF, Greece;
BSF-NSF and GIF, Israel;
Norwegian Financial Mechanism 2014-2021, Norway;
NCN and NAWA, Poland;
La Caixa Banking Foundation, CERCA Programme Generalitat de Catalunya and PROMETEO and GenT Programmes Generalitat Valenciana, Spain;
G\"{o}ran Gustafssons Stiftelse, Sweden;
The Royal Society and Leverhulme Trust, United Kingdom.
 
The crucial computing support from all WLCG partners is acknowledged gratefully, in particular from CERN, the ATLAS Tier-1 facilities at TRIUMF (Canada), NDGF (Denmark, Norway, Sweden), CC-IN2P3 (France), KIT/GridKA (Germany), INFN-CNAF (Italy), NL-T1 (Netherlands), PIC (Spain), ASGC (Taiwan), RAL (UK) and BNL (USA), the Tier-2 facilities worldwide and large non-WLCG resource providers. Major contributors of computing resources are listed in Ref.~\cite{ATL-SOFT-PUB-2021-003}.
 

\printbibliography

\clearpage 
 
\begin{flushleft}
\hypersetup{urlcolor=black}
{\Large The ATLAS Collaboration}

\bigskip

\AtlasOrcid[0000-0002-6665-4934]{G.~Aad}$^\textrm{\scriptsize 99}$,    
\AtlasOrcid[0000-0002-5888-2734]{B.~Abbott}$^\textrm{\scriptsize 125}$,    
\AtlasOrcid[0000-0002-7248-3203]{D.C.~Abbott}$^\textrm{\scriptsize 100}$,    
\AtlasOrcid[0000-0002-2788-3822]{A.~Abed~Abud}$^\textrm{\scriptsize 34}$,    
\AtlasOrcid[0000-0002-1002-1652]{K.~Abeling}$^\textrm{\scriptsize 51}$,    
\AtlasOrcid[0000-0002-2987-4006]{D.K.~Abhayasinghe}$^\textrm{\scriptsize 91}$,    
\AtlasOrcid[0000-0002-8496-9294]{S.H.~Abidi}$^\textrm{\scriptsize 27}$,    
\AtlasOrcid[0000-0002-9987-2292]{A.~Aboulhorma}$^\textrm{\scriptsize 33e}$,    
\AtlasOrcid[0000-0001-5329-6640]{H.~Abramowicz}$^\textrm{\scriptsize 158}$,    
\AtlasOrcid[0000-0002-1599-2896]{H.~Abreu}$^\textrm{\scriptsize 157}$
% ...
\bigskip
\\

% ...
$^{27}$Physics Department, Brookhaven National Laboratory, Upton NY; United States of America.\\
% ...
$^{33}$$^{(a)}$Facult\'e des Sciences Ain Chock, R\'eseau Universitaire de Physique des Hautes Energies - Universit\'e Hassan II, Casablanca;$^{(b)}$Facult\'{e} des Sciences, Universit\'{e} Ibn-Tofail, K\'{e}nitra;$^{(c)}$Facult\'e des Sciences Semlalia, Universit\'e Cadi Ayyad, LPHEA-Marrakech;$^{(d)}$LPMR, Facult\'e des Sciences, Universit\'e Mohamed Premier, Oujda;$^{(e)}$Facult\'e des sciences, Universit\'e Mohammed V, Rabat;$^{(f)}$Mohammed VI Polytechnic University, Ben Guerir; Morocco.\\
$^{34}$CERN, Geneva; Switzerland.\\
% ...
$^{51}$II. Physikalisches Institut, Georg-August-Universit\"{a}t G\"ottingen, G\"ottingen; Germany.\\
% ...
$^{91}$Department of Physics, Royal Holloway University of London, Egham; United Kingdom.\\
% ...
$^{99}$CPPM, Aix-Marseille Universit\'e, CNRS/IN2P3, Marseille; France.\\
$^{100}$Department of Physics, University of Massachusetts, Amherst MA; United States of America.\\
% ...
$^{125}$Homer L. Dodge Department of Physics and Astronomy, University of Oklahoma, Norman OK; United States of America.\\
% ...
$^{157}$Department of Physics, Technion, Israel Institute of Technology, Haifa; Israel.\\
$^{158}$Raymond and Beverly Sackler School of Physics and Astronomy, Tel Aviv University, Tel Aviv; Israel.\\
% ...
$^{a}$ Also at Borough of Manhattan Community College, City University of New York, New York NY; United States of America.\\
$^{b}$ Also at Bruno Kessler Foundation, Trento; Italy.\\
$^{c}$ Also at Center for High Energy Physics, Peking University; China.\\
$^{d}$ Also at Centro Studi e Ricerche Enrico Fermi; Italy.\\
$^{e}$ Also at CERN, Geneva; Switzerland.\\
$^{f}$ Also at D\'epartement de Physique Nucl\'eaire et Corpusculaire, Universit\'e de Gen\`eve, Gen\`eve; Switzerland.\\
% ...

\end{flushleft}


\end{document}